\documentclass[%
 reprint,
nofootinbib,
 amsmath,amssymb,
 aps,
prd,
]{revtex4-2}

\usepackage[hidelinks,colorlinks=true,linkcolor=blue,citecolor=blue]{hyperref}
\usepackage{natbib}
\usepackage{CJK}
\usepackage{graphicx}% Include figure files
\usepackage{dcolumn}% Align table columns on decimal point
\usepackage{bm}% bold math
\def\eg{\textit{e.g.}}

\begin{document}

%\preprint{APS/123-QED}

\title{A novel category of environmental effect in gravitational waves \\ from binaries perturbed by periodic forces}% Force line breaks with \\
%\thanks{A footnote to the article title}%

\author{Lorenz Zwick$^{1}$}
\email{lorenz.zwick@nbi.ku.dk}
 %\altaffiliation[Also at ]{Physics Department, XYZ University.}%Lines break automatically or can be forced with \\
\author{Christopher Tiede$^{1}$}
\author{Alessandro A. Trani$^{1}$}
\author{Andrea Derdzinski$^{2,3}$}
\author{Zoltan Haiman$^{4,5}$}
\author{Daniel J. D'Orazio$^{1}$}
\author{Johan Samsing$^{1}$}
\affiliation{%
 $^{1}$Niels Bohr International Academy, The Niels Bohr Institute, Blegdamsvej 17, DK-2100, Copenhagen, Denmark\\
$^{2}${Department of Life and Physical Sciences, Fisk University, 1000 17th Avenue N., Nashville, TN 37208, USA} \\
$^{3}${Department of Physics \& Astronomy, Vanderbilt University,
2301 Vanderbilt Place, Nashville, TN 37235, USA}\\
$^{4}$Department of Astronomy, Columbia University, 550 West 120th Street, New York, NY 10027, USA\\
$^{5}$Department of Physics, Columbia University, 550 West 120th Street, New York, NY 10027, USA
}%

%\collaboration{MUSO Collaboration}%\noaffiliation

%

%\collaboration{CLEO Collaboration}%\noaffiliation

\date{\today}% It is always \today, today,
             %  but any date may be explicitly specified

\begin{abstract}
We study the gravitational wave (GW) emission of sources perturbed by periodic dynamical forces which do not cause secular evolution in the orbital elements. We construct a corresponding post-Newtonian waveform model and provide estimates for the detectability of the resulting GW phase perturbations, for both space-based and future ground-based detectors. We validate our results by performing a set of Bayesian parameter recovery experiments with post-Newtonian waveforms. We find that, in stark contrast to the more commonly studied secular dephasing, periodic phase perturbations do not suffer from degeneracies with any of the tested vacuum binary parameters. We discuss the applications of our findings to a range of possible astrophysical scenarios, finding that such periodic perturbations may be detectable for massive black hole binaries embedded in circum-binary discs, extreme mass-ratio inspirals in accretion discs, as well as stellar-mass compact objects perturbed by tidal fields. We argue that modelling conservative sub-orbital dynamics opens up a promising new avenue to detect environmental effects in binary sources of GWs that should be included in state-of-the-art waveform templates.
%\begin{description}
%\item[Usage]
%Secondary publications and information retrieval purposes.
%\item[Structure]
%You may use the \texttt{description} environment to structure your abstract;
%use the optional argument of the \verb+\item+ command to give the category of each item. 
%\end{description}
\end{abstract}

%\keywords{Suggested keywords}%Use showkeys class option if keyword
                              %display desired
\maketitle

%\tableofcontents
\section{Introduction}\label{sec:introduction}
Astrophysical sources of gravitational waves (GW) are likely to form in high density environments, such as star clusters, nuclear clusters, and accretion or circumbinary discs. Understanding and modelling the physical mechanisms at play in these environments is crucial, as they are responsible for determining the efficiency of the various formation channels for black hole binaries (BHB), and ultimately result in the expected event rates of detectable signals for current and future GW detectors \citep[see the following reviews and white papers, among the innumerable relevant works,][]{2010abadie,2013dominik,2021mapelli,2022lisaastro, DOrazioCharisi:2023,2024lisa}. In addition to determining the rates, different astrophysical environments set characteristic ranges and correlations in the expected parameters of observed BHBs, as exemplified by the numerous efforts to link LIGO/Virgo/Kagra observations to specific formation channels \citep{2002belczynski,2007oleary,2008Sadowski,2016antonini,2017vitale,kavanagh2020,2021zevin,2021kimball,2023santini}.

In almost all astrophysically motivated scenarios, the direct influence of the source's environment on the emitted GWs is expected to be extremely small once BHBs  reach the separations required to produce detectable GWs \citep[with the possible exception of Pulsar Timing Array sources, see, \eg,][]{2009haiman,2013sesana}. However, extremely small need not necessarily mean negligible. These influences are referred to with the umbrella term of "environmental effects" (EE), which are defined as small perturbations with respect to the expected vacuum GWs that arise as a consequence of the source's astrophysical environment. Common examples of EEs include dynamical perturbations caused by third bodies, effects of peculiar velocity and other accelerations, as well as the study of frictional forces motivated by both gaseous and dark matter backgrounds \citep{2008barausse,inayoshi2017,2017meiron,2017Bonetti,2019alejandro,2019randall,DOrazioGWLens:2020, 2022liu,2022xuan,garg2022,2022cole,2022chandramouli,2023zwick,2023Tiede,2024dyson}.

Since the early mentions in the nineties \citep{1993chakrabarti,1995ryan} to the seminal work of the mid two-thousands \citep{2007levin,kocsis,2014barausse}, the study of EE, i.e. astrophysically motivated perturbations to GW emission, has been gaining traction due to the ever closer dawn of space-based GW detectors sensitive to the milli-Hz band \citep[in particular the recently adopted LISA mission,][]{2024lisa}. The scientific motivations are threefold: Firstly, the detection of even a single source that is clearly perturbed by EEs would allow to set unprecedented constraints on the surroundings of BHBs and the physical processes that facilitate their formation and shape their evolution \citep{2019andrea,2021andrea,2022zwick,2022speri,2022cole,2022destounis,garg2022,2024samsing}. Secondly, the presence of unmodelled EEs may produce biases in the recovered binary parameters (see references above). Thirdly, deviations from the expected vacuum evolution of sources may be mislabelled as resulting from non-standard physics, in the case that astrophysical influences are not modelled correctly \citep[see, \eg,][]{2017moore,2019yamada,2020cardosomaselli,2020cardosoee,2022cole,2023zwick}. The latter point is especially relevant when interpreting signals from extreme mass-ratio inspirals, which are supposed to provide extremely precise measurement of the metric in the close vicinity of massive BHs \citep{2007barack,2013gair,2017lisa,2022polcar}. In short, the careful study of EE is required to both preserve and broaden the scientific goals of future GW detectors in the joint fields of fundamental physics and astrophysics.

Focusing now on binary sources of GWs, the state of the art in the modelling of EE currently consists of the implementation of simple secular de-phasing prescriptions in pre-existing vacuum waveform templates, as well as their study through Bayesian parameter estimation techniques \citep[][]{2021toubiana,2022cole,2022speri,2024garg}. Indeed, the observable that is most often associated to EE is that of a phase difference between the perturbed waveform and a reference vacuum waveform, which slowly accumulates throughout the binary source's inspiral phase. The main motivations for focusing on de-phasing are twofold: Firstly, the process of matched filtering is known to be extremely sensitive to cumulative perturbations to the GW phase. Phase shifts $\delta \phi$ as small as:
\begin{align}
    \label{eq:dephcritbad}
    \delta \phi \sim \text{few}/\text{SNR},
\end{align}
where SNR is the signal to noise ratio of the given GW, can significantly reduce the match between a waveform template and the data and reduce the effectiveness of parameter estimation \citep[see, \eg,][]{1994cutler,2008damour,kocsis,2021katz,owen23}. Secondly, most environmental effects may indeed be modelled with simple secular de-phasing prescriptions, at least as a first approximation \citep[see][for an impression on the broad relevance of dephasing in the context of EE]{2023zwick}. As an example, any coupling between the environment's and the binary's energy will cause a de-phasing by the following relation:
\begin{align}
\delta \phi \sim \int \int \left( \Delta \dot{f} \, {\rm{d}}t\right) \, \, {\rm{d}}t,
\end{align}
where  $\Delta \dot{f}$ is a change in the binary's chirp due to the coupling. Additionally, any kind of EE that involves a Doppler shift \citep[see, \eg,][where modifications to the GW phase and GW amplitude are discussed in detail]{2019chamberlain}, such as peculiar motion or gravitational red-shift, is also well approximated via a dephasing \citep{2017meiron,inayoshi2017,2019alejandro,2021chen,2023yan}:
\begin{align}
  \delta \phi \sim \int \Delta f  \, {\rm{d}}t,
\end{align}
where $\Delta f$ is the Doppler shift in frequency. 

Secular de-phasing is often considered the most promising avenue to detect EE in GWs, and several works have used the reference value of Eq. \ref{eq:dephcritbad} to claim detectability\footnote{Including, admittedly, the corresponding author's own work.}. While the latter criterion can provide a good estimate in certain limits, several careful parameter inference studies have highlighted how de-phasing suffers from many-fold intrinsic limitations which may greatly hinder the prospective of correctly identifying it in realistic binary sources \citep[][]{2020cardoso,2023Keijriwal,owen23}. Firstly, secular dephasing is preferentially accumulated in the very early inspiral stages of a binary's observable evolution, where the binary only chirps slowly. This corresponds to a rather low available signal-to-noise (SNR) in which the perturbation may effectively be constrained. Secondly, tracking an accumulated phase shift over a very long inspiral may bee exceedingly hard due to the intrinsic challenges of matched-filtering \citep{2021ewing}. Thirdly, secular de-phasing is similar to many other post-Newtonian (PN) corrections to the binary's vacuum evolution, which also add slowly accumulating contributions to the total GW phase \citep[see \eg ][]{blanchet2014}. Finally, and perhaps most crucially, it is the case that a hypothetical constant dephasing term, i.e. a phase offset, has exactly the same form as the source's initial phase (as can be seen at a glance in, \eg, Eq. \ref{eq:wavemod2}), leading to potential degeneracies. The consequence of this is that only the rate of variation in the de-phasing is truly informative, rather than the magnitude of the de-phasing itself.

These intrinsic challenges raise the following questions: Are there any other features in GWs from perturbed binaries that may carry detailed information regarding the source's interaction with its environment? Are any of these features less prone to the inherent limitations of secular dephasing when it comes to parameter estimation? And is there any expectation that GW detectors may be able to detect these in the not too distant future? 
Motivated by these questions, we take a step away from modelling forces or perturbations that would produce any long term evolution on the binary's orbital elements, since such forces ultimately appear in the GW as various flavours of secular dephasing. Instead, we focus entirely on forces that may influence the binary's motion on orbital or even sub-orbital timescales without causing any long term evolution. We will find that such forces leave characteristic and potentially extremely informative imprints in the GW emission of binaries. \newline \newline

This work is structured as follows:
In section \ref{sec:GWfrombin}, we solve the equations of motion for a circular binary that is influenced by an arbitrary, periodic perturbing force. We then use the latter to construct a ready-to-use post Newtonian waveform model. In sections \ref{sec:detmethods} and \ref{sec:detectability}, we analyse the resulting periodic perturbations in the GW phase and provide some detectability criteria using both analytical and numerical tools. In section \ref{sec:astro} we discuss a range of plausible astrophysical scenarios in which such perturbations may be detectable. Finally, we lay out concluding remarks in section \ref{sec:conclusion}.

\begin{figure}
    \centering
    \includegraphics[width=0.95\columnwidth]{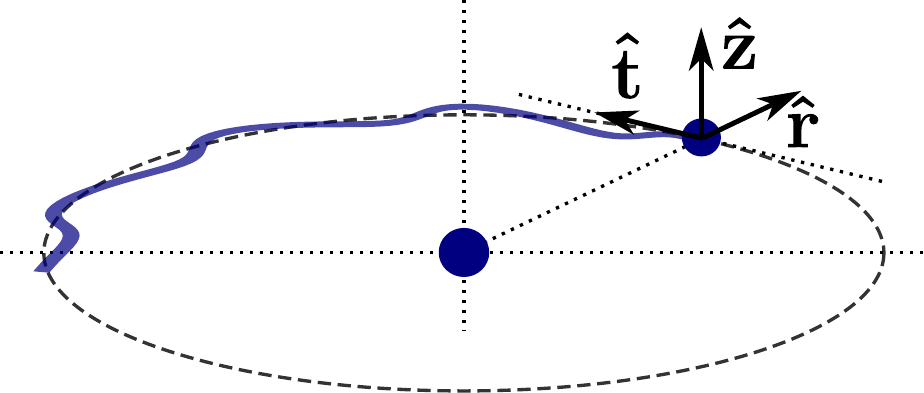}
    \includegraphics[width=\columnwidth]{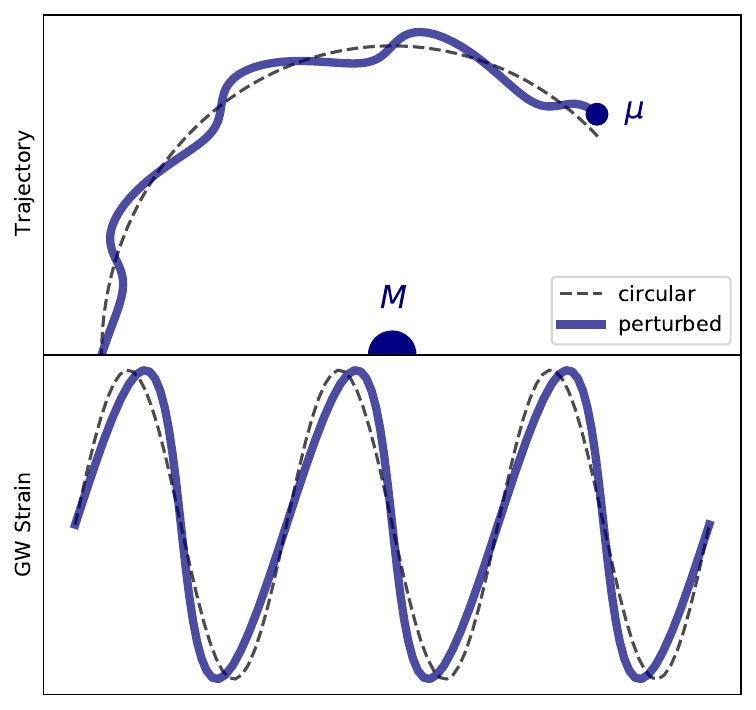}
    \caption{A simple representation of the type of perturbations analysed in this work. The sketch showcases a binary (represented by a total mass $M$ and reduced mass $\mu$) being affected by time-varying forces (decomposed in their tangential, radial and normal components). Here we only consider forces that do not cause a secular evolution in the orbital elements. The latter influence the binary's instantaneous velocity and lever arm, which in turn modifies the system's GW emission. The GW strain will therefore present features associated to the forces acting on the binary. The magnitude of the effects are greatly exaggerated for clarity.}
    \label{fig:sketch}
\end{figure}

\section{Gravitational waves from binaries subject to time-varying forces}
\label{sec:GWfrombin}
\subsection{Motion of a circular binary under time varying forces}
We consider the case of a binary source of GWs with arbitrary total mass $M$ and reduced mass $\mu$. The binary's orbit is described by the solutions to the six osculating elements $\left[p,e,\nu,\omega,\Omega, I \right]$\footnote{Semi-latus rectum, eccentricity, true anomaly, argument of peri-apsis, longitude of ascending node and inclination.}, which fix the coordinate position of the reduced mass in space \citep[following here the formalism detailed in][]{2014poisson}: 
\begin{align}
    \begin{pmatrix}
        x \\ y \\ z
    \end{pmatrix} = 
    r \begin{pmatrix}
        \cos\Omega\cos\left(\omega + \nu\right) - \sin \Omega \sin \left(\omega + \nu \right) \cos I \\
        \sin \Omega \cos \left(\omega + \nu\right) +\cos \Omega \sin \left(\omega + \nu \right) \cos I \\ \sin\left(\omega + \nu\right) \sin I
    \end{pmatrix},
\end{align}
where the instantaneous radius is:
\begin{align}
    r = \frac{p}{1 + e \cos \nu},
\end{align}
In the GW driven regime, most sources will be circularised by the effect of radiation reaction. Therefore, we will operate in the limit of $e \to 0$ as a first approximation, and comment on the possible effects of eccentricity below.

We study the effect of a perturbing force ${\rm{d}} \mathbf{F}$ which is acting on the binary, which represents the influence of the environment with respect to the vacuum gravitational force. Here ${\rm{d}} \mathbf{F}$ has units of acceleration, and may be decomposed into its radial, tangential and normal components:
\begin{align}
    {\rm{d}} \mathbf{F} = R \mathbf{\hat{r}} + T \mathbf{\hat{t}} + N \mathbf{\hat{z}},
\end{align}
where $\mathbf{\hat{r}}$, $\mathbf{\hat{t}}$ and $\mathbf{\hat{z}}$ denote unit vectors in the aforementioned directions (see Fig. \ref{fig:sketch}).

In this work, we restrict our analysis to in-plane perturbative forces, as out of plane displacements will only ever affect the orbital inclination angles. Note that, if sufficiently strong, variations in the inclination angles may well produce detectable modulations in the relative amplitude of the GW's plus and cross polarisations. We leave this effect for future investigation. We treat the binary's orbit in the Newtonian limit. While this assumption is known to break down as the orbital velocity approaches relativistic values, it is justifiable by two arguments: First, most EEs will affect the binary more strongly at large separations, where orbital speeds are low and GW emission is weaker. Second, the assumption is justified in a strictly perturbative sense, as additional cross terms\footnote{Cross terms between environmental perturbations and post Newtonian dynamics have been extensively studied in \cite{2014will,2014will2}, and have interesting consequences on the secular evolution of triple systems and orbits in accretion discs \citep{2023zwickflare}.} between the environmental perturbations and relativistic dynamics will be suppressed with respect to the leading order terms in $1/c^2$. {However, see \cite{2022cardoso} for a fully relativistic approach appropriate for high mass-ratio binaries}. Note that under these assumptions, the binary's motion will be confined to a unique plane, which we identify as the $(x,y)$ plane.

In the Newtonian limit, the evolution equations for the osculating orbital elements are known from the seminal Lagrange planetary equations:
\begin{align}
    \frac{{\rm d}p}{{\rm d}t} = 2 \sqrt{\frac{p^3}{G M}} \frac{T}{\left(1 + e \cos \nu \right)}
\end{align}
\begin{align}
    \frac{{\rm d}\nu}{{\rm d}t} &= \sqrt{\frac{G M}{p^3}}(1+ e \cos \nu )^2\nonumber \\ &+ \frac{1}{e}\sqrt{\frac{p}{GM}}\left[ \cos \nu R - \frac{2+ e \cos \nu}{1+ e \cos f}\sin \nu T\right] 
\end{align}
\begin{align}
    \frac{{\rm d}e}{{\rm d}t} = \sqrt{\frac{p}{GM}}\left[R \sin \nu + \frac{2 \cos \nu + e(1+\cos^2 \nu)}{1+e \cos \nu}T \right]
\end{align}

\begin{align}
     \frac{{\rm d}\omega}{{\rm d}t} &= \frac{1}{e}\sqrt{\frac{p}{GM}}\Big[ - \cos \nu R + \frac{2+ e \cos \nu}{1+ e \cos \nu}\sin \nu \,T\Big]
\end{align}
which we can solve in the perturbative limit via standard techniques. Our goal is to capture the effect of forces which do not cause a secular evolution in the orbital elements of the binary. Therefore, we the follow the groundwork laid out in \cite{2022zwick} and describe a generic force as a sum of Fourier components with a fundamental frequency equal to the binary's orbital frequency $f_{\rm K}$:
\begin{align}
    \label{eq:perturbingforces}
   {\rm{d}} \mathbf{F}^{\rm p} &= \text{Re}\Bigg(\sum_n \Big[ B^{\rm{T}}_n(r) \exp\left(i2 \pi   n f_{\rm K} t\right) \mathbf{\hat{t}} \nonumber \\ &+ B^{\rm{R}}_n(r) \exp\left(i2 \pi  n f_{\rm K} t\right) \mathbf{\hat{r}}\Big]\Bigg) ,
\end{align}
%JS: how can this form describe say gas friction?
where the complex coefficients $B^{\rm T}_n$ and $B^{\rm R}_n$ are only functions of the binary's separation and their magnitude and dependence on $r$ encodes the forces that are applied on the binary by the given EE. The form of Eqs. \ref{eq:perturbingforces} is to be understood as a Fourier series in the limit in which the binary residency time at the separation $r$ is much larger than one orbital period, i.e. the adiabatic limit. In that case, we can also allow (with a slight abuse of notation) the parameter $n$ to take non-integer values in order to describe periodic perturbations that do not exactly match a multiple of the system's dynamical time. We can insert the perturbing force into the Lagrange planetary equations to find analytical solutions to the orbital elements. As it will be directly related to GW observables, we report the explicit first-order perturbative solution\footnote{Here we take the chance to note an imprecision present in \cite{2022zwick}, in which the computations did not properly propagate the perturbation in the Kepler equation and resulted in a mistake in Eq. (33). However, the latter does not change the qualitative results in \cite{2022zwick}, as the physical scaling remains unaffected.} of the true longitude $\theta = \nu + \omega$:
\begin{align}
    \label{eq:thetasol}
    \theta &= 2 \pi f_{\rm K}t + \sum_n \left[ \delta \theta_n ^{\rm T}(t) + \delta \theta_n ^{\rm R}(t )\right]  \\
    \delta \theta_n^{\rm T} &= \frac{B_n^{\rm T}p_0^2 \exp(i 2\pi n f_{\rm K}t)}{G M \left(n^6-5 n^4+4n^2\right)}\Big(-12 + 7n^2-n^4  \nonumber \\ &+ 2n^2(2+n^2) \mathcal{C}_2 -i6n^3\mathcal{S}_2) \Big)    \\
    \delta \theta_n^{\rm R} &=-\frac{B_n^{\rm R}p_0^2 \exp(i 2\pi n f_{\rm K}t)}{G M \left(n^6-5 n^4+4n^2\right)}n\Big(i(-4 + n^2)  \nonumber \\ &+ i3n^2 \mathcal{C}_2 -n(2+n^2)\mathcal{S}_2) \Big)
\end{align}
where $p_0$ is the semi-latus rectum of the unperturbed orbit, $\mathcal{S}_m \equiv \sin(2\pi m f_{\rm K})$ and $\mathcal{C}_m \equiv \cos(2\pi m f_{\rm K})$, and only the real part of the perturbation is relevant. Here the results are shown for the case $n \neq [0,1,2]$, which are discussed below.

\begin{figure}
    \centering
    \includegraphics[width=\columnwidth]{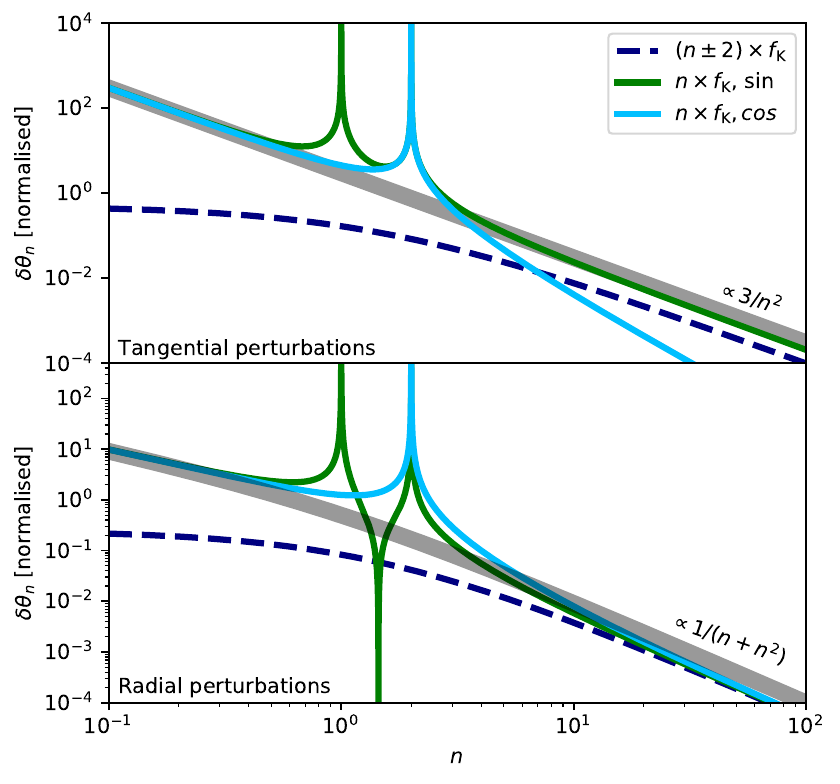}
    \caption{We plot the normalised amplitudes of the true longitude perturbation coefficients (Eqs. \ref{eq:thetasol}). Both radial and tangential perturbations are split into several components, at frequencies of $\left[n-2,n,n+2 \right]f_{\rm K}$ and in sinusoidal and co-sinusoidal parts. The dominant perturbations are tangential at the frequency $n f_{\rm K}$, and their scaling with the parameter $n$ can be well approximated by $3/n^2$ when disregarding resonances. We do not fit the $n\pm2$ perturbations (dashed lines) as they are sub-dominant. Note that tangential perturbations always dominate unless $n>>1$.}
    \label{fig:coefficients}
\end{figure}

The remaining orbital elements take the schematic form:
\begin{align}
    p(t) &= p_0 + \delta p(t) \\
    e(t) &=   \delta e(t) \\
    \omega(t) &=  \delta \omega(t) \\
    \nu(t) &= 2 \pi f_{\rm K}t + \delta \nu(t),
\end{align}
where the perturbations (denoted by $\delta$) are purely oscillatory {and vanish when averaged over the true anomaly on timescales longer than the forcing (again for $n \neq [0,1,2]$}). We also report the explicit solution for the binary's separation, which oscillates about its unperturbed value:
\begin{align}
    \label{eq:radius}
    r(t) &=p_0 + \sum_n\Bigg[\frac{B^{\rm R}_n p_0^3\exp(i 2\pi n f_{\rm K}t)}{GM(1-n^2)} \nonumber \\  &- \frac{i 2B^{\rm T}_n p_0^3\exp(i 2\pi n f_{\rm K}t)}{GMn(1-n^2)}\Bigg].
\end{align}
We find, as expected, that oscillatory forces cause a corresponding behaviour in the orbital elements with a periodicity comparable to the forcing. A visualisation of the resulting binary motion can be seen in Figure \ref{fig:sketch}. In essence, Eqs. \ref{eq:thetasol} and \ref{eq:radius} state that a periodic, in-plane perturbation with an amplitude $B$ and with a periodicity of $n f_{\rm K}$ will produce variations in the orbital phase of typical order $B p_0^2/(GM)$ at the different harmonics $(n-2,n, n+2)$. The amplitude of high-frequency fluctuations is suppressed as $1/n^2$, due to the shorter available time to apply an acceleration. We can summarise the results schematically as follows:
\begin{center}
\begin{tabular}{c c c}

    Force & $\to$ & Phase \\
  $\big\{B,[n ]f_{\rm K}\big\}$ & $\to$ & $\Big\{ \sim \frac{Bp_0^2}{G M},[n-2,n,n+2 ] f_{\rm K} \Big\}$ ,
\end{tabular}
\end{center}
{where the central perturbation corresponds to the direct effect of tangential or radial forces on the binary phase, while the two other arise as a response of the binary's elements to the changing instantaneous velocity or radial separation, respectively.} In Fig. \ref{fig:coefficients} we show the dependence of the true longitude perturbation coefficients on the periodicity $n$. We find that, when neglecting some special values of $n$ as discussed below, the coefficients for tangential and radial perturbations can be well fit by the simpler expressions $3/n^2$ and $1/(n+n^2)$, respectively. {Note that these approximations are valid both in the limit $n\to 0$ and $n\to \infty $, and will be used in the numerical tests of section \ref{sec:detectability}.} \newline \newline

The cases of $n=0$, $n=1$, and $n=2$ must be treated separately in order to avoid divergences. They can be understood as the binary's orbital frequency entering into resonance with a corresponding component of the perturbing force. In general, resonances are an important factor in determining the evolution of perturbed binaries, and may even determine the dynamical stability of triple systems \citep{chirikov1979,mardl2008}. Additional PN contributions to the binary's motion, such as the general relativistic apsidal precession of the binary, may also resonate with the outer forcing, either constructively or destructively. Finally, we further note that the interaction of periodic forces with eccentricity will give rise to further resonances at higher epi-cyclical harmonics. For the remainder of this analysis, we will not model the effect of resonances, leaving them for future investigation.

\subsection{Vacuum and perturbed waveform models}
The GW emission of a binary depends on the physical parameters describing its position in space, i.e. its radius $r$, the true longitude $\theta$ as well as the inclination angles with respect to the observer. In the Newtonian limit, the relation between the source and the gravitational radiation far away is given by the quadrupole formula:
\begin{align}
    \label{eq:quadrupole}
    h_{ij}(t) = \frac{2G}{c^4 D} \Ddot{\mathcal{I}}_{ij}(t_{\rm r}),
\end{align}
where $\mathcal{I}$ is the mass quadrupole tensor in its traceless form, $D$ is the source's luminosity distance and $t_{\rm r}$ is the retarded time. In our case, additional perturbations to the quadrupole tensor arise from from both the radius and the true anomaly (Eqs. \ref{eq:thetasol} and \ref{eq:radius}). We insert the explicit solutions to the orbital elements in Eq. \ref{eq:quadrupole} and proceed to calculate the components of the GW strain. After some intensive but straightforward algebraic computations we find that the perturbed gravitational waveform can be expressed in the following way:
\begin{align}
    \label{eq:gwschematic}
    h \sim h_{\rm N}(\theta') + \delta h(\delta r,\delta \dot{r},\delta \ddot{r},\delta \dot{\theta},\delta \ddot{\theta}),
\end{align}
where we defined the perturbed variable $\theta' = \theta + \delta \theta$ as in Eq. \ref{eq:thetasol}. Here $h_{\rm N}$ is the familiar Newtonian quadrupolar waveform, while $\delta h$ are additional perturbations to the radiated GW. The latter have been studied in detail in \cite{2022zwick}, and only contain terms explicitly proportional to $\delta r$, $\delta \dot{r}$, $\delta \ddot{r}$, $\delta \dot{\theta}$ and $\delta \ddot{\theta}$. The interpretation of this result is that the perturbed waveform consists of two separate parts: First, a Newtonian waveform evaluated at a perturbed phase. Second, a series of additional low-amplitude harmonics dependent on the perturbations in the orbital elements and their derivatives, which arise directly from variations in the binary's lever arm $r + \delta r$ or radial and tangential radial velocities.
\newline \newline
%JS: are other effects competing with Eq. 21, e.g.,
%doppler phase shift, or what about GW redshift along the orbit as the binary component wiggels back and forth? 
All the components in Eq. \ref{eq:gwschematic} carry information regarding the forces acting on the binary and could in principle be included in a fully general perturbed waveform model. Additionally, the mass distributions responsible for producing the forces could themselves produce additional GW radiation. For the purposes of this work, we focus solely on the first part of Eq. \ref{eq:gwschematic}, led by the standard intuition that phase perturbations are more easily detectable than amplitude perturbations of the same order \citep{1994cutler}. Furthermore, we will see that perturbations of this type already contain an incredible amount of information regarding the forces acting on the binary. Our goal is thus to construct a waveform model that can account for such periodic phase perturbations over the full observation of a chirping source. We base our methodology on the seminal work in \cite{1994cutler} and construct a Fourier space GW template using the stationary phase approximation, which is justified as long as the perturbations are sufficiently small. In the stationary phase approximation, a Newtonian waveform with a phase $\phi_{\rm GW}$ is given by:
\begin{align}
    \Tilde{h}(f) = \frac{Q}{D}\left(\frac{G\mathcal{M}}{c^3}\right)^{5/6}f^{-7/6} \exp \left[ i (- \phi_{\rm GW} - \pi/4 ) \right],
\end{align}
where $f$ is the observed GW frequency, $\mathcal{M}$ is the red-shifted chirp mass, $D$ is the luminosity distance, $Q$ is a geometric pre-factor that accounts for projections of the wave onto a given detector and the tilde denotes a Fourier transform. In our case then, the GW phase is simply the sum of the vacuum phase and the phase perturbations:
\begin{align}
    \phi_{\rm GW} &= \phi_{\rm vac} + 2\sum_n \left[ \delta \theta_n ^{\rm T}(t) + \delta \theta_n ^{\rm R}(t )\right] \\  \phi_{\rm vac} &=  \phi_{\rm c}- 2 \pi f t  +  2\left(8\pi \mathcal{M}f \right)^{-5/3} (+\text{ PN corrections}),
\end{align}
where $\phi_c$ is the initial phase and the factor 2 multiplying the sum arises from the relation $\phi_{\rm GW}=2\theta$. Note that now it is important to make explicit the radial dependence of the force Fourier coefficients also in terms of the GW frequency, due to the frequency dependence of Eqs. \ref{eq:thetasol} and \ref{eq:radius}:
\begin{align}
    B^{\rm{T}}_n &\to B^{\rm{T}}_n(p_0) \text{ or } B^{\rm{T}}_n(f)\\
    B^{\rm{R}}_n &\to B^{\rm{R}}_n(p_0) \text{ or } B^{\rm{R}}_n(f).
\end{align}
Higher accuracy in terms of post-Newtonian terms may also easily be achieved by applying the perturbed phase appropriately to the required higher order GW modes \footnote{One may also easily add PN corrections to the binary's phase without adding the corresponding higher-order GW multipole terms, as is commonly done in analytical waveform models}. Adapting the notation in \cite{2014blanchet} we have:
\begin{align}
    \label{eq:wavemod}
    h(f) = \frac{Q}{D}\left(\frac{G\mathcal{M}}{c^3}\right)^{5/6}f^{-7/6}\times\sum_{lm} \mathcal{H}^{lm} \exp\left[- im\psi\right],
\end{align}
where the mode coefficients $\mathcal{H}^{lm}$ are known to very high post-Newtonian orders, and:
\begin{align}
    \label{eq:wavemod2}
   \psi &= -\pi/4 + \phi_{\rm c}- 2 \pi f t  +  2\left(8\pi \mathcal{M}f \right)^{-5/3} \nonumber \\ &+\text{ PN corrections} + 2\sum_n \left[ \delta \theta_n ^{\rm T}(t) + \delta \theta_n ^{\rm R}(t )\right].
\end{align}
Eqs. \ref{eq:wavemod} and \ref{eq:wavemod2} represent a post-Newtonian, quasi-circular waveform model that can describe the effect of arbitrary time-varying forces on the binary's phase. The number of additional free parameters with respect to a vacuum waveform is limited to the number of free parameters required to specify the perturbations on the binary. For clarity's sake, let us summarise again the assumptions that went into the construction of the waveform model:
\begin{itemize}
    \item We focus on circular binaries subject to in-plane perturbative forces.
    \item The interaction between the relativistic binary and the perturbative forces is treated at lowest order, i.e. Newtonian.
    \item Any secular evolution of the orbital elements caused by perturbative forces is removed by construction, including that arising from resonances.
    \item Perturbations to the GW emission are similarly treated at lowest order and only phase perturbations are considered. The additional low amplitude harmonics were studied in \cite{2022zwick}.
    \item GW emission resulting from the stress energy tensor of the environment causing the perturbation is neglected.
    
\end{itemize}
Given these assumptions, we propose our perturbed waveform model as a proxy for more sophisticated templates that can account for both strong gravity effects and environmental perturbations in a fully relativistic way.
We highlight the exciting prospect of measuring forces directly from, \eg, hydrodynamical and GRMHD simulations of gas-embedded binaries or precise integrations of few body dynamical scatterings, including them in Eqs. \ref{eq:wavemod} and \ref{eq:wavemod2}, and performing parameter estimation directly on such numerical results.

\section{SNR criteria}
\label{sec:detmethods}
\subsection{Baseline model and Newtonian SNR estimates.}
\label{sec:essmodel}
From this point on, we will omit to specify both the summation and the distinction between radial and tangential forces, and simply refer to a generic phase perturbation caused by a force of magnitude $B$ with a periodicity of frequency $n f_{\rm K}$. From Eqs. \ref{eq:thetasol}, we know that the phase perturbation will consist of three distinct oscillations at the characteristic frequencies $[2-n,n,2+n]f_{\rm K}$, and a characteristic dimensionless amplitude$ \sim B_n p_0^2/(G M)$.
We can further simplify the waveform model presented in Section \ref{sec:GWfrombin} by assuming that the central component of the phase perturbation triplet $[2-n,n,2+n]f_{\rm K}$ dominates, as suggested by Fig. \ref{fig:coefficients}. Furthermore, we assume that the radial dependence of the Fourier coefficients follows a power law.

Under these assumptions, we can then construct a simplified version of the resulting GW-phase perturbation:
\begin{align}
    \label{eq:ess}
     \phi^{\rm ess}(f) &= \phi_{\rm{vac}}(f) + B' \left(\frac{f}{f_{\rm{in}}}\right)^{\beta'}  \cos(n f t(f) + \phi_0^{\rm B} ),
\end{align}
where $B'$ is a dimensionless amplitude which we scale at the frequency $f_{\rm in}$ at which the source becomes observable to a particular detector. Eq. \ref{eq:ess} still captures the qualitative effect of the complete true longitude perturbations on the actual GW phase, and is more easily treated for order of magnitude estimates. We therefore refer to it as a baseline model for the perturbations treated in this work. Note that, for clarity's sake, here and from now on, the primed quantities are always specifically related to GWs, while the non-primed quantities relate to the forces acting on the binary. If the radial dependence of the perturbing force follows a power law with an exponent $\beta$, we obtain the following relations between the forces and the phase perturbations:
\begin{align}
\label{eq:bprime}
     \beta' &= \beta - 4/3, \\
    B' &\approx\frac{6 B_n}{n^2(G M)^{1/3}}\pi^{-4/3}f_{\rm in}^{-4/3},
\end{align}
where in the last step we replace the polynomial in $n$ from Eqs. \ref{eq:thetasol} with the simplified fit of $3/n^2$ (see also Fig. \ref{fig:coefficients}), and here $\beta'$ describes the frequency dependance.

For the purposes of this work, we will compute simple SNR estimates using the standard definition of the noise-weighted scalar product for two waveforms $h_1$ and $h_2$:
\begin{align}
    (h_1,h_2) = 2\int_0^{\infty} \frac{\Tilde{h}_1^{*}\Tilde{h}_2 + \Tilde{h}_1\Tilde{h}_2^{*}}{S_n} {\rm{d}}f',
\end{align}
where $S_n$ is the detector's noise power spectral density. The signal to noise of a GW is then defined simply as:
\begin{align}
    \text{SNR}^2=(h,h).
\end{align}
We also  use a simple prescription to assess the detectability of GW phase perturbations, which is commonly used in works that treat environmental effects. We require the noise-weighted scalar product of the GW perturbation to exceed a certain threshold which we leave unspecified, but is commonly chosen to be $\sim \mathcal{O}(10)$:
\begin{align}
    \label{eq:deltasnr}
    \delta\text{SNR}^2=(\delta h,\delta h) = 8 \int \frac{\Tilde{h}^2}{S_n}\left[ 1 - \cos \left(\delta \phi \right)\right]\, {\rm{d}} f'.
\end{align}
where the prefactor 8 arises from the normalisation of the one sided power spectrum \citep{2019robson}. Note that the latter equation is often only a rough estimate of the detectability of a waveform perturbation. In particular, the commonly chosen criterion $\delta$SNR$>$8 does not account for degeneracies and correlations between the binary vacuum parameters and the perturbations under scrutiny. Thus, Eq. \ref{eq:deltasnr} often overestimates the detectability of secular de-phasing, which is strongly degenerate with both the binary's initial phase as well as other parameters such as effective spin\footnote{This particular degeneracy is in fact responsible for the mismatch between the Bayesian analysis performed in this work and recent works \citep[see, \eg,][]{2024garg} on gas de-phasing based on the \texttt{pycbc} package, as the latter treats the initial phase as an \textit{extrinsic} parameter.}, as seen in Fig. \ref{fig:mcmc2}. For this reason, we will test the simple detectability criteria with several MCMC parameter recovery tests.  From here on, we will refer to the results of Eq. \ref{eq:deltasnr} as a "$\delta$SNR" estimate.
\newline \newline

Let us also briefly discuss secular de-phasing. In binaries perturbed by external forces, secular de-phasing is caused by the exchange of energy and angular momentum between the binary and a separate reservoir. It is mediated by a residual tangential force $A_0^{\rm T}(r)$, from which we remove any time variation by performing a time average. We can find the resulting secular drift in binary separation by using the Lagrange planetary equations:
\begin{align}
    \dot{p} = 2A_0^{\rm T}(p_0)\sqrt{\frac{p_0^3}{GM}}.
\end{align}
When it comes to GW observables, the additional drift in orbital separation modifies the source's chirp evolution:
\begin{align}
    \frac{\dot{p}}{p} = -\frac{3}{2}\frac{\dot{f}}{f} =-\frac{3}{2}\frac{\dot{f}_{\rm{GW}}+\dot{f}_{\rm A}}{f},
\end{align}
where:
\begin{align}
    \dot{f}_{\rm GW} &= \frac{96}{5} \pi ^{8/3}\left(\frac{G \mathcal{M}}{c^3}\right)^{5/3}f^{11/3},\\
    \dot{f}_{\rm A} &= - \frac{4}{3\pi} A_0^{\rm T}(f)\left(\frac{f^2} {G M\pi^2}\right)^{1/3}.
\end{align}
The accumulated GW phase over the binary chirp is then given by:
\begin{align}
    \phi =2 \pi \int \frac{f}{\dot{f}_{\rm{GW}}+\dot{f}_{\rm gas}} \, {\rm{d}}f \approx \phi_{\rm vac} - \int \frac{f \dot{f}_{\rm A}}{\dot{f}_{\rm{GW}}^2}{\rm{d}}f,
\end{align}
where:
\begin{align}
    \phi_{\rm vac} \approx \frac{1}{32 \pi^{8/3}}\left(\frac{G \mathcal{M}}{c^3}\right)^{-5/3}f_{\rm in}^{-5/3}
\end{align}
The cumulative effect of a tangential force thus causes a secular de-phasing of the form:
\begin{align}
    \phi^{\rm d}(f) &= \phi_{\rm{vac}}(f) + A'\left(\frac{f}{f_{\rm{in}}}\right)^{\alpha'},
\end{align}
where $A'$ is a dimensionless coefficient and we assumed that the radial dependence of the secular force follows a power law with an exponent $\alpha$. The general relation between the force and the secular de-phasing is:
\begin{align}
    \alpha' &= \alpha -14/3 \\
    A'&= \frac{A_0^{\rm T}}{(G M)^{1/3}}\frac{25f_{\rm in}^{-14/3}}{2304(14-3\alpha) \pi ^7}\left(\frac{G  \mathcal{M}}{c^3} \right)^{-10/3}.
\end{align}
Comparing this result with Eq. \ref{eq:bprime}, we find that the amplitude of secular de-phasing will typically accumulate in the early inspiral and decay extremely rapidly as the source chirps.
\subsection{Bayesian parameter recovery with post-Newtonian waveforms.}
In this paper we perform several numerical parameter recovery tests to validate the simpler analytical prescriptions of the previous sections. We adopt a vacuum waveform model of the type described by Eqs. \ref{eq:wavemod} and \ref{eq:wavemod2}. We include the phase evolution of a non-spinning binary up to third PN order by using the formulas from \cite{2014blanchet}. Similarly, we also account for all higher order modes beyond quadrupolar GW emission up to order $x^{5/2}$, where $x$ is the post-Newtonian parameter. Following \cite{1994cutler}, we also include the leading order effect of spin through the effective parameter $\mathcal{B}$, which enters the phase at first-and-a-half post-Newtonian order and represents a projection of the orbital angular momentum and the individual binary component spins. In this work, we do not include any spin evolution term nor the effects of spin-spin coupling. Finally, we include the effect of periodic dephasing by means of the baseline model detailed above, for a single periodicity. While this suffices to obtain correct detectability estimates, we stress that the latter cannot replace the full formulas for true longitude perturbations in actual applications.

We limit our analysis to the binary's intrinsic parameters and initial phase, while performing a sky average of a source's position on the sky and orientation. Furthermore, we assume that the binary coalescence time may be measured with high precision through the merger and ring-down portions of the GW signal, which we do not model here. Therefore, our post-Newtonian waveform is completely specified by the following nine independent parameters:
\begin{align}
\label{eq:paramssss}
    \text{Binary} = &\begin{cases}
        z; & \text{Redshift}\\
        \mathcal{M}; & \text{Chirp mass}\\
        \mu; & \text{Reduced mass}\\
        \phi_{\rm c}; & \text{Initial binary phase}\\
        \mathcal{B}; & \text{Effective spin}
    \end{cases}\\
    \text{Perturbation} =&\begin{cases}
        \log_{10}(B'); & \text{Initial amplitude}\\
        \beta'; & \text{Frequency scaling}\\
        n; & \text{Periodicity}\\
        \phi_0^{B}; & \text{Init. perturbation phase}  
    \end{cases}      
\end{align}
We use Monte-Carlo-Markhov-Chain  (MCMC) methods to perform parameter estimation studies numerically.
The goal of MCMC based parameter estimation is to sample the posterior distributions of a set of waveform parameters $\Theta$ given a likelihood function $\mathcal{L}$ of the form:
\begin{align}
    \mathcal{L}\left(\Theta \right) \propto \exp\left[ - \big(h(\Theta) - h(\Theta_{\rm GT}),h(\Theta) - h(\Theta_{\rm GT}) \big) \right]
\end{align}
where $\Theta_{\rm GT}$ is a vector of parameters for the injected signal. The detector noise is treated as being stationary, representing the average value for the noise power spectral density over many realisations. We use the affine invariant sampler \texttt{emcee} \citep{2013emcee} to perform the numerical tests, typically running 32 parallel walkers for approximately 30'000 to 50'000 steps, where the typical auto-correlation time of the walkers is $\sim 200$. We initialise the walkers in the vicinity of the injected true values in order to speed up convergence and to avoid having to specify priors for the vacuum binary parameters (which is appropriate only for high SNR sources). The priors for the perturbation parameters are simply chosen for convenience and do not influence our results unless stated otherwise.
\section{Detectability of Periodic Phase perturbations}
\label{sec:detectability}
\subsection{Detectability for monochromatic sources}
\label{sec:mono}
\begin{figure}
    \centering
\includegraphics[width=\columnwidth]{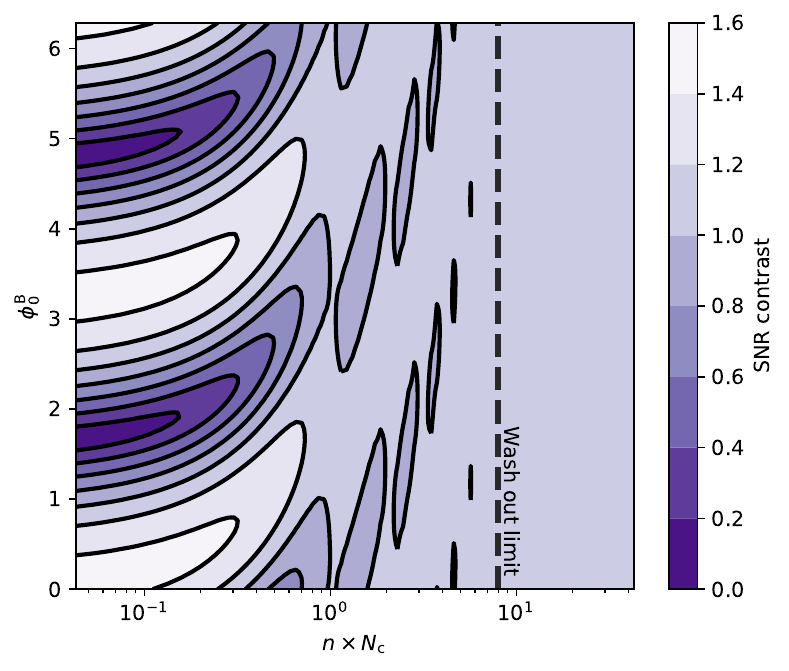}
    \caption{We show the variations in the $\delta$SNR caused by varying the parameters $n$ and $\phi_0^{\rm B}$ in Eq. \ref{eq:wwww} for a typical chirping source that completes $N_{\rm c}$ cycles in band. As expected, we find that chirping sources that complete many cycles tend to wash out the importance of the perturbation's initial phase. The latter's variation is completely negligible as long as the source typically completes more than $\sim 10/n$ cycles at each corresponding frequency bin.}
    \label{fig:wash}
\end{figure}

Here we use the baseline GW phase model constructed in Sec \ref{sec:essmodel} to extract some qualitative insight regarding the detectability of periodic perturbations. We start from the definition of $\delta$SNR:
\begin{align}
    \label{eq:monosnr}
     \delta\text{SNR}^2=(\delta h,\delta h) = 8 \int \frac{\Tilde{h}_{\rm vac}^2}{S_n}\left[ 1 - \cos \left(\delta \phi \right)\right]\, {\rm{d}} f',
\end{align}
and want to evaluate it for:
\begin{align}
\label{eq:wwww}
   \delta \phi = B' \left(\frac{f}{f_{\rm{in}}}\right)^{\beta'}  \cos(n f t(f) + \phi_0^{\rm B} ),
\end{align}
representing a periodic perturbation with a typical amplitude $B'$ at the GW frequency $f_{\rm in}$, a frequency scaling given by $\beta'$, a typical periodicity described by the frequency $n f_{\rm K}$ and an inital phase $\phi_0^{\rm B}$. Evaluating Eq. \ref{eq:monosnr} in the limit of small $B'$ and also for monochromatic sources, results in the following expression:
\begin{align}
    \label{eq:deltasnrperiodic}
    \delta\text{SNR}^2 &\approx \frac{\Tilde{h}_{\rm vac}^2\dot{f}}{S_{\rm n}} B'^2\left(\frac{f}{f_{\rm{in}}}\right)^{2 \beta'} \int^{T} \cos(n f t + \phi_0^{\rm B})^2\, {\rm d}t.
\end{align}
where we replaced the integration in frequency with an integration over the observation time $T$. Here we can already see that, as $T$ is much longer than a wave period, the dependence on the parameters $n$ and $\phi_{0}^{\rm B}$ should be washed away by the integration. This is also true for chirping sources, as long as the number of cycles at a given frequency is larger than $\sim n^{-1}$, as shown in Fig. \ref{fig:wash}. The average of the integral then simply evaluates to $\sim 1/2 \times T$ and reduces to the convenient form:
\begin{align}
    \label{eq:quadrature}
   \delta\text{SNR}^2 \approx \frac{1}{2}\text{SNR}(f)^2 B'(f)^2,
\end{align}
where we used the definition of the signal-to-noise ratio for a monochromatic unperturbed source \citep[see, \eg,][]{2014blanchet,2018maggiore,2019robson}. We can also easily extrapolate the results of performing the integral in Eq. \ref{eq:deltasnrperiodic} for a full spectrum of perturbations. Re-introducing the subscript $n$ to denote the various periodic force components, we find that the total SNR grows in quadrature:
\begin{align}
    \delta{\text{SNR}}_{\rm tot} = \text{SNR} \times \sqrt{\frac{1}{2} \sum_{n}(B'_{n})^2} \approx \text{SNR} \times B' \sqrt{N/2 } ,
\end{align}
where in the latter approximation we define $N$ as an effective number of periodic perturbations with amplitude $B'$. The latter factor can significantly boost the detectability of periodic force perturbations from EE that naturally produce perturbations at many frequencies, such as, \eg, the complex gas hydrodynamics around merging binaries, see Sec. \ref{sec:astro}.
Finally, we report the threshold acceleration required for detection of a single periodic perturbation:
\begin{align}
    B_n^{\rm det} &> A_{\rm G} \frac{n^2\sqrt{2}}{3 \times \text{SNR}} \\ &=  \frac{n^2}{\text{SNR}}\frac{\pi^{4/3} \sqrt{2}}{6}(G M)^{1/3}f^{4/3},
\end{align}
where $A_{\rm G}$ is the binary's gravitational acceleration. This equation states that individual periodic force components are detectable if they are of the order of $A_{\rm G}/n^2/$SNR, which strongly prefers sub-orbital perturbations. 
\subsubsection{Comparison with secular dephasing}
Eq. \ref{eq:deltasnrperiodic} suggests that for monochromatic sources, phase oscillations of approximately $\delta \phi \sim {\text{few }} \times 1/$SNR should be detectable, which seems comparable to the analogous requirement for secular de-phasing, in the limit where degeneracies may be ignored. The comparison is however not trivial, as in the latter case dephasing arises as a consequence of the slow, coherent accumulation of the perturbation over the entire observation. For almost monochromatic sources, the GW phase accumulates as:
\begin{align}
    \phi &= \int_0^{T}\left[ f + \dot{f}(f)t \right]\, \rm{d}t.
\end{align}
where $T$ is the observation time. In case of a secular perturbation, the total accumulated dephasing $\Delta \phi$ is given by:
\begin{align}
    \Delta \phi =\frac{1}{2}\dot{f}_{\rm A}T^2
\end{align}
Evaluating Eq. \ref{eq:deltasnr}, the expected $\delta$SNR produced by secular de-phasing amounts to:
 \begin{align}
    \delta\text{SNR}^2 &\approx \frac{\Tilde{h}_{\rm vac}^2\dot{f}}{S_{\rm n}} \dot{f}_{\rm gas}^2\int^{T} t^4\, {\rm d}t = \frac{4}{5}\text{SNR}(f)^2\Delta \phi^2.
\end{align}
The previous equation speaks to the high potential detectability of secular dephasing. It suggests that, indeed, {\textit{cumulative}} shifts of order few $ 1/\text{SNR}$ may be detectable. However, this conclusion ignores the fact that secular dephasing, by its own nature of being slowly accumulated, can be degenerate with many binary parameters that act in a similar way. In particular, it is highly degenerate with the binary's initial phase, along with any post-Newtonian modifications to the binary's chirp behaviour with a similar magnitude or frequency scaling. A more conservative requirement for the detectability of secular dephasing is that it should exceed a full cycle, i.e. $2 \pi$, while many works have also highlighted cases where several cycles of dephasing are necessary to perform parameter estimation \citep[see, \eg,][]{2022speri,2023Keijriwal}. Here we use $2 \pi$ as a detectability criterion, finding that the threshold acceleration for secular de-phasing amounts to:
\begin{align}
    A_0^{^{\rm det}} = 3 \pi^{5/3} (G M)^{1/3} f^{-2/3} T^{2}.
\end{align}
The ratio between the detection thresholds for periodic and secular dephasing yields some insight:
\begin{align}
   \frac{B_n^{^{\rm det}}}{A_0^{^{\rm det}}} = \frac{1}{{\text{SNR}}}\frac{\sqrt{2}4n^2 }{9\pi ^{1/3}}(f T)^{2} \sim \frac{N_{\rm c}^2}{\text{SNR}},
\end{align}
where $N_{\rm{c}}$ is the number of cycles at the frequency $f$. Interestingly, this suggests that periodic perturbations may be the dominant EE for high SNR sources that complete comparatively fewer cycles in the sensitivity band of their respective detectors. As we will see in Sec. \ref{sec:astro}, this may be realised in several astrophysical scenarios. The equation above also suggests that, in sources where both secular and periodic forces are present, the latter must be significantly higher in amplitude to be detected at the same $\delta$SNR. However a more precise statement on detectability requires a careful Bayesian analysis. Indeed, as we will see in \ref{sec:degen}, these naive expectations are often misleading. 

\subsection{Detectability for chirping LISA sources}
When stepping beyond the monochromatic case, it is necessary to account for the full noise power spectral density of a detector. Here we produce some simple detectability criteria for the recently adopted, space based gravitational wave detector LISA \citep{2022lisaastro,2024lisa}, but note that similar estimates apply to any similar milli-Hz GW instrument \citep[\eg\ TianQin][]{2021tian}.

\begin{figure}
    \centering
\includegraphics[width=\columnwidth]{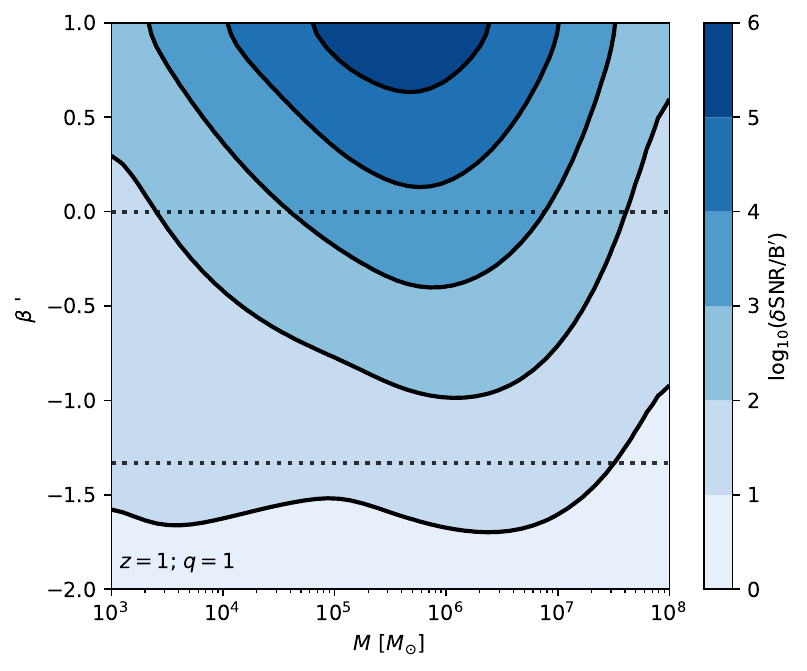}
\vspace{0cm}
\includegraphics[width=\columnwidth]{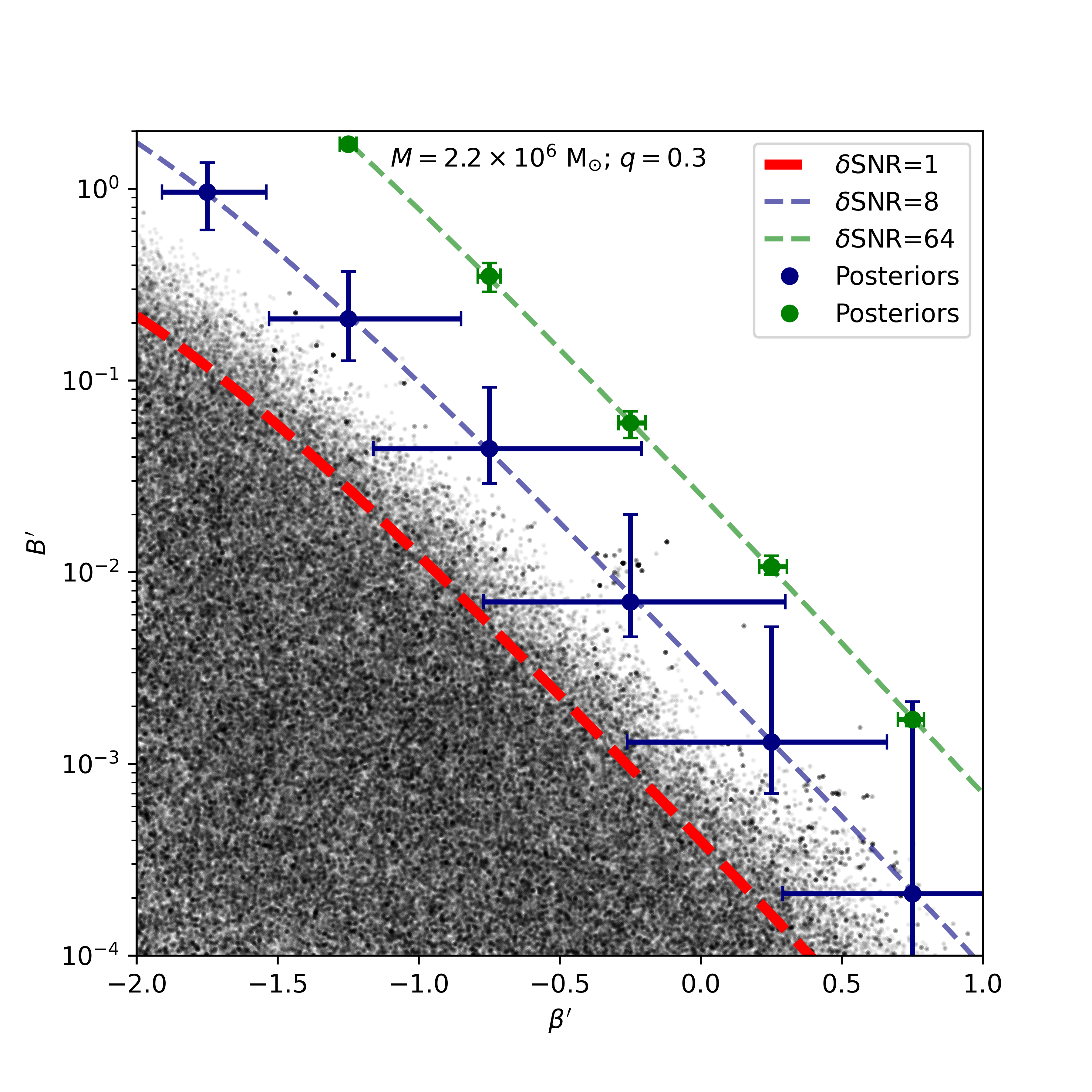}
    \caption{Top panel: The detectability contours for a periodic dephasing with amplitude $B'$ and frequency scaling $\beta'$, for a wide range of LISA sources with total mass $M$, mass-ratio $q=1$ at redhsift $z=1$. The dashed lines highlight the cases of a constant perturbation amplitude ($\beta'=0$) or a constant force ($\beta'=-4/3$) throughout the chirp. Bottom panel: The recovered marginalised posteriors of $\beta'$ and $B'$ from several MCMC parameter estimation tests, with injected perturbations that lie along the $\delta$SNR = 8 (blue) and $\delta$SNR = 64 (green) threshold lines. The grey dots represent the density of samples of an MCMC run testing the "null hypothesis", i.e. the absence of a perturbation. The latter is excluded at $1\sigma$ for perturbations with $\delta$SNR above 1, which is represented by the red dashed line. }
    \label{fig:lisadet}
\end{figure}

As a proxy for LISA's noise power spectral density, we use the convenient fit from \cite{2019robson} and evaluate Eq. \ref{eq:deltasnr} for a range of perturbation parameters $B'$ and $\beta'$. We repeat the process for an array of source total masses, mass-ratios and redshifts. The $\delta$SNR results for the baseline perturbations, for typical massive BHB sources, are visualised as a contour plot in the top panel of Fig. \ref{fig:lisadet}. The $\delta$SNR contours have a strong dependence on the parameter $\beta'$, and generally align with the expectations set from the monochromatic case, i.e. they scale inversely with the source's total SNR and linearly with the perturbation amplitude $B'$. Here we highlight some results for the case of a constant amplitude for the phase perturbation, $\beta'= 0$, and the case for a constant amplitude in radius or frequency of the perturbing force, $\beta' = -4/3$. For the former, the perturbations are detectable for a wide range of sources whenever $B' \sim 10^{-3}$ to $10^{-2}$. For the latter, detectability requires $B' \sim 0.1$. The following formulae provide a simple fit to the results of Fig.~\ref{fig:lisadet} at $z=1$:
\begin{align}
    \frac{\delta {\text{SNR}}}{B'} \approx \sqrt{\frac{2\mu}{M}}\begin{cases}
     \left( 10^{1.2\beta' + 2.5} + 5.9 \right) & {\text{for }} M=10^4\, \rm{M}_{\odot}\\
    \left( 10^{1.8\beta' + 3.5} + 3.0 \right) & {\text{for }} M=10^5\, \rm{M}_{\odot}
    \\  \left( 10^{1.8\beta' + 3.8} + 3.0 \right) & {\text{for }} M=10^6\, \rm{M}_{\odot}
    \\ \left( 10^{1.2\beta' + 2.9} + 1.0 \right) & {\text{for }} M=10^7\, \rm{M}_{\odot}
    \end{cases}
\end{align}
where we separated typical LISA binary mergers into several mass decades.

We confirm these estimates with a series of numerical tests, also partially shown in the bottom panel of Fig. \ref{fig:lisadet}. We take a representative source with $M = 2.2 \times 10^6$ M$_{\odot}$ and define three detectability threshold lines at $\delta$SNR=1, 8 and 64 according to Eq. \ref{eq:deltasnr}.
We then perform several MCMC parameter recovery experiments with post-Newtonian waveforms, in which all parameters detailed in Eq. \ref{eq:paramssss} are sampled freely. We plot the standard deviations of the marginalised posterior distributions for $\beta'$ and $B'$  as errorbars in the bottom panel of Fig. \ref{fig:lisadet} for a series of injected values of $B'$ and $\beta'$ that lie along the corresponding $\delta$SNR lines. We find that all injected perturbations can be recovered, provided that they lie above a $\delta$SNR of a few. In general we find the following qualitative correspondence between the $\delta$SNR and the MCMC based parameter recovery:
\begin{align}
    &\text{$\delta$SNR} &\text{MCMC Results} \nonumber \\
    & \sim 1 &\implies\text{"Null hypothesis" excluded at 1$\sigma$.} \nonumber \\
     & \sim 8 &\implies\text{1$\sigma$ uncertainty $<$ 0.5  for $\log_{10}(B')$ and $\beta'$}. \nonumber \\
     & \sim 64 &\implies\text{1$\sigma$ uncertainty $<$ 0.05  for $\log_{10}(B')$ and $\beta'$} \nonumber.
\end{align}
Here the "null hypothesis" represents the complete absence of a perturbation within the waveform. We have tested the results showcased in Fig.~\ref{fig:lisadet} for a range of total masses between $M= 10^3$ M$_{\odot}$ and $M = 10^8$ M$_{\odot}$, with mass-ratios from $q=0.1$ to $q=1$ and for a few redshifts between $z=0.5$ and $z=3$. Since the aforementioned qualitative trends seem to hold without fail, we conclude that our $\delta$SNR based detection criteria are robust, and can provide a good estimate for detectability even when accounting for parameter degeneracies and post-Newtonian modelling for the unperturbed vacuum waveforms. We emphasise that this is in stark contrast to secular de-phasing, as further discussed below.

\subsection{Detectability for future ground based detectors.}
Here we perform the same analysis as above, while adopting the prospective sensitivity curve of the Einstein Telescope (ET) as a proxy for the next generation of ground based GW detectors \citep{2020maggioreet}. The implications of the latter on the astrophysics of stellar-mass black holes will be further discussed in Sec. \ref{sec:astro}.

\begin{figure}
    \centering
\includegraphics[width=\columnwidth]{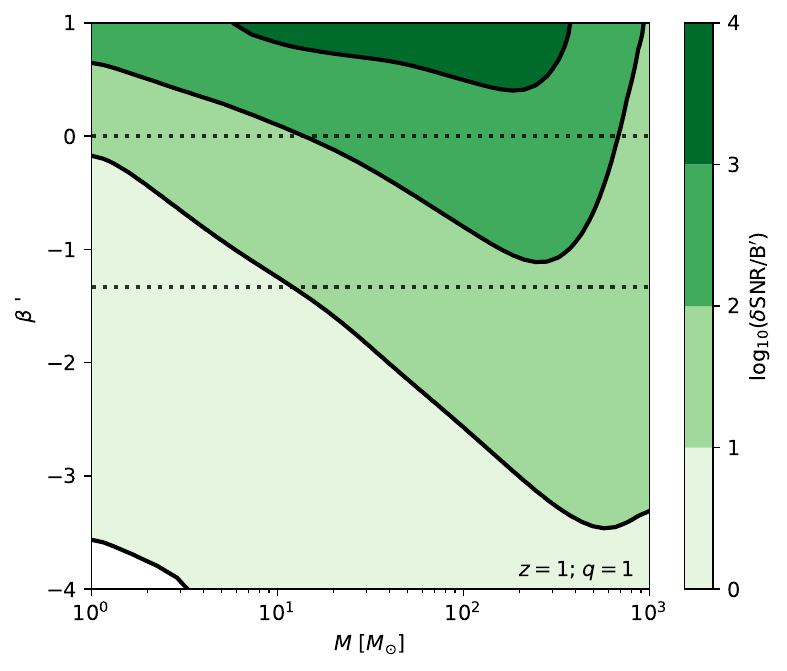}
    \caption{Same as Fig. \ref{fig:lisadet}, using the noise power spectral density of ET \citep{2020maggioreet} as a proxy for future ground based detectors.}
    \label{fig:cedet}
\end{figure}

The results of our analysis are showcased in Fig. \ref{fig:cedet}, and similar considerations apply here as in the previous section. The highest detectability is achieved for sources with total masses of $\sim$few$\times 10^2$ M$_{\odot}$, which maximise the SNR of the detection. This is comparable to the highest observed masses in the current LIGO catalogue of sources, GW190521 \citep{2020ligobig}. Interestingly, detecting EE in latter type of sources may also be possible with LISA, during the very early inspiral stages \citep{SesanaMultiband:2016, GerosaMultiband:2019,2021toubiana,2022sberna}. We highlight that with the given ET sensitivity, the detectability of periodic perturbations in chirping sources is plausible for a much larger range of $\beta'$ than for LISA sources, reaching down to $\beta' \sim -3$ for high SNR sources, due to the different form of the noise power spectral density. Similarly to the previous section, we provide fits to the resulting $\delta$SNR, which may be used as a simple estimate:
\begin{align}
    \frac{\delta {\text{SNR}}}{B'} \approx \sqrt{\frac{2\mu}{M}}\begin{cases}
     \left( 10^{1.4\beta' + 1.4} + \beta' + 5 \right) & {\text{for }} M=3\, \rm{M}_{\odot}\\
    \left( 10^{0.8\beta' + 2.2} + 1.0 \right) & {\text{for }} M=30\, \rm{M}_{\odot}
    \\  \left( 10^{0.6\beta' + 2.1} + 1.7 \right) & {\text{for }} M=300\, \rm{M}_{\odot}
    \end{cases}
\end{align}

\subsection{Test case: Degeneracies in secular vs periodic dephasing}
\label{sec:degen}

\begin{figure*}
    \centering
    \includegraphics[width=2\columnwidth]{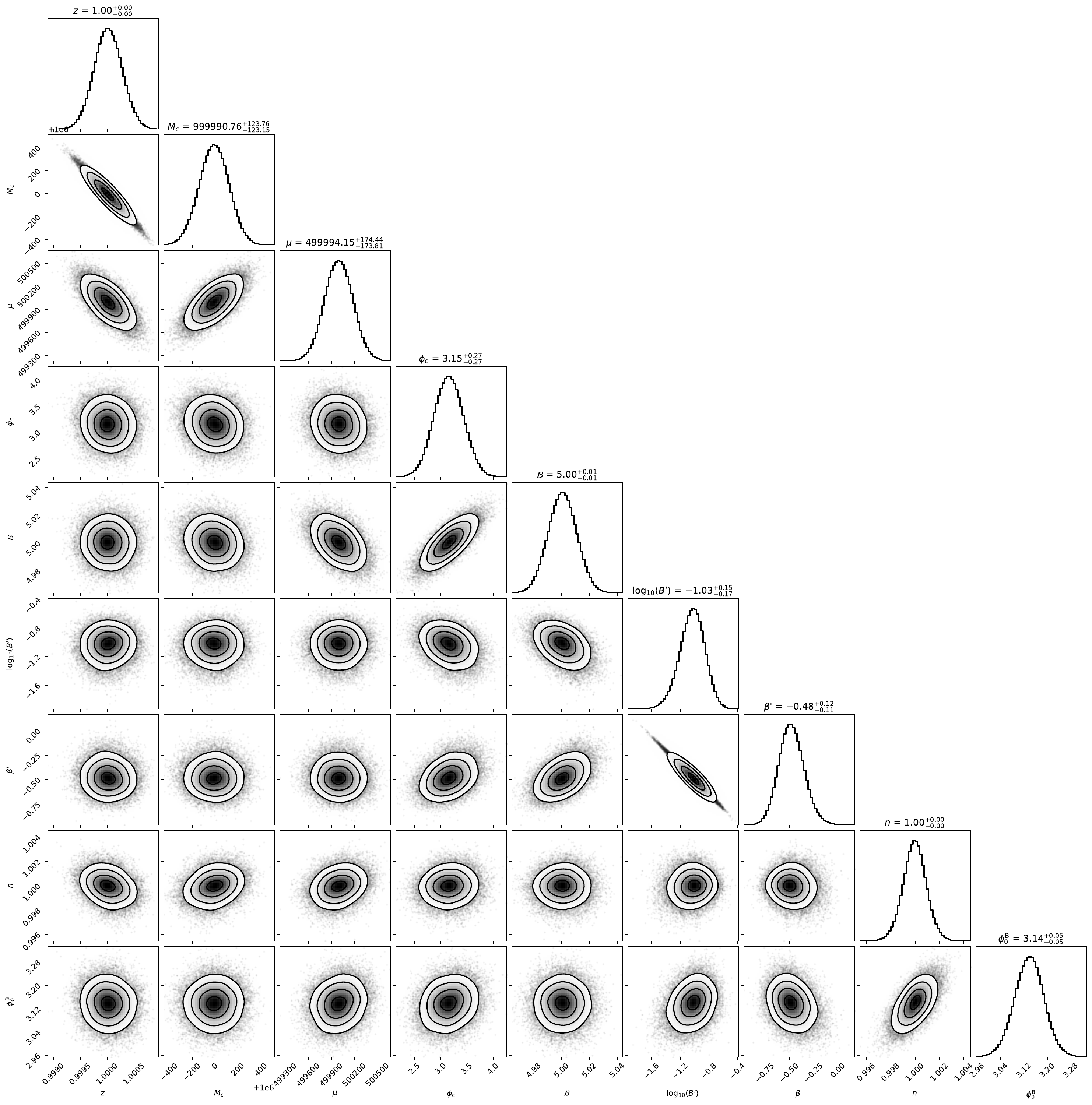}
    \caption{Full showcase of the marginalised posteriors of a high SNR post-Newtonian waveform. The injected periodic dephasing has a Newtonian $\delta$SNR of approximately 40, and can be recovered very precisely. Note that the frequency of the perturbation can be recovered with a precision of 0.1$\%$. Crucially, no correlations between the periodic perturbation and the vacuum binary parameters appear. The MCMC is performed with 32 parallel walkers for 50'000 steps, requiring approximately 45 minutes of computation on a single core. See Eqs. \ref{eq:paramssss} for the meaning of the nine parameters.}
    \label{fig:mcmc1}
\end{figure*}

In Fig \ref{fig:mcmc1}, we showcase the full 9 parameter posteriors (in black) from a realisation of an MCMC parameter estimation test for a representative high-SNR LISA source, where we injected a strong periodic de-phasing with $B'=0.1$, $n=1$ and $\beta' = -0.5 $, resulting in a $\delta$SNR of approximately 40. All of the two dimensional marginalised parameter posteriors are well behaved, clearly localised and centered on the true value. The only correlations visible in the binary parameters are limited to the familiar chirp mass and redshift degeneracy, as well as the initial phase and effective spin parameter degeneracy. The only degeneracy that appears in the recovered perturbation is between the amplitude $B'$ and the frequency scaling $\beta'$. Investigating the latter degeneracy in more detail, we find that by defining two new parameters:
\begin{align}
    \left[ \beta' , \log_{10}(B') \right] \to \left[ \kappa_1 , \kappa_2 \right]
\end{align}
where:
\begin{align}
    \beta' =- \kappa_1 \times  \log_{10}(B') + \kappa_2
\end{align}
one can completely characterise the correlation in $B'$ and $\beta'$. Here we do not complete a full analysis of this, but note that by performing the parameter estimation on $\kappa_1$ and $\kappa_2$, one can determine the latter within approximately $1\sigma$ of the true injected value already for sources with $\delta$SNR $\sim 1$. Therefore, while the parameters $\kappa_1$ and $\kappa_2$ are less informative than $B'$ and $\beta$, they can be used to distinguish the presence of a perturbation from its absence, and are therefore connected to the red dashed line shown in Fig. \ref{fig:lisadet} and described in the text.

The test case showcased in Fig. \ref{fig:mcmc1} is representative of all $\delta$SNR larger than a few, where the widths of the Gaussian posteriors roughly scale as $1/\delta$SNR. This suggests that the additional parameters required to describe periodic de-phasing are completely orthogonal to the usual vacuum binary parameters. Crucially, they do not cause any visible degradation in the recovery of the vacuum waveform. {Note that here we used a value of $n = 1$ as a proxy for any perturbation on close to orbital timescales, still neglecting the secular effect of resonances. The results for the posteriors with diferent values of n roughly follow the expectations given by the $n^{-2}$ scaling of Eq. \ref{eq:ess}. }

\begin{figure*}
    \centering
\includegraphics[width=2\columnwidth]{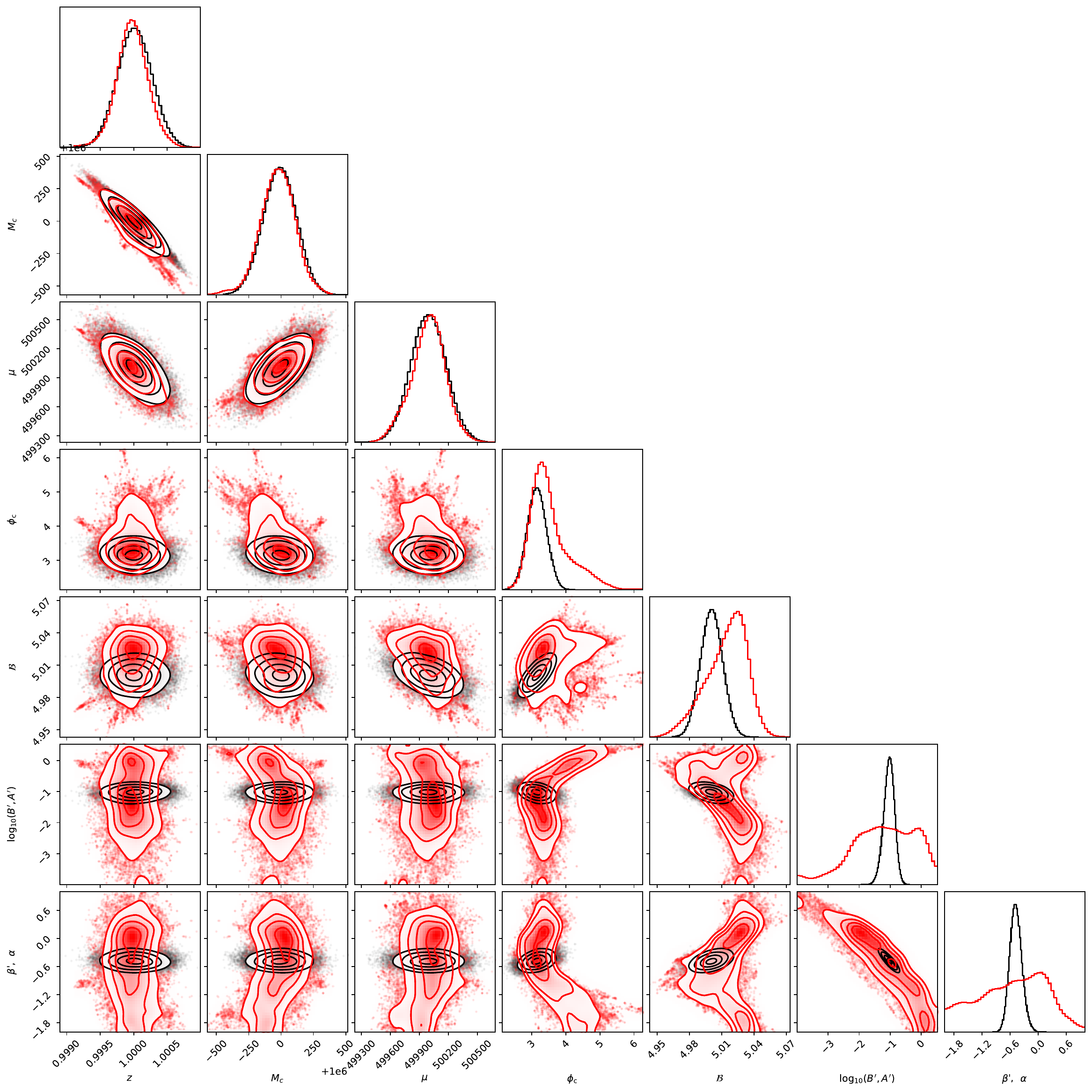}

    \caption{Full showcase of the marginalised posteriors when attempting to recover a periodic (black) and a secular (red) dephasing. The injected periodic dephasing is characterised by $B' =0.1$ and $\beta$ has a $\delta$SNR of approximately 40. the injected secular dephasing has the same amplitude and scaling, resulting in a $\delta$SNR of approximately 60. Nevertheless, the latter cannot be reconstructed confidently as it suffers from a degeneracy with the binary's initial phase and spin parameter. The two MCMCs are performed with 32 parallel walkers for 50'000 steps, requiring approximately 90 minutes of computation on a single core. See Eqs. \ref{eq:paramssss} for the meaning of the nine parameters.}
    \label{fig:mcmc2}
\end{figure*}

We compare the aforementioned result with an attempt to recover the same vacuum binary source while adding a secular de-phasing term, with a corresponding amplitude $A'=0.1$ and frequency scaling $\alpha'=-0.5$. The latter perturbation actually corresponds to significantly higher $\delta$SNR with respect to the comparative case for a periodic perturbation, i.e. approximately 60 instead of 40 (according to Eq. \ref{eq:deltasnr}). In Fig. \ref{fig:mcmc2}, we showcase the difference in the recovered posteriors between the two cases. 

We highlight some interesting differences: For the injected value of $B' = A' = 0.1$, the recovered value of the periodic perturbation is extremely well constrained, yielding $B'_{\rm rec}= 0.094 \pm 0.04$ at $1\sigma$. In contrast, the posterior for the amplitude of the secular perturbation is highly degenerate and essentially completely unconstrained, allowing for values that range all the way from $\sim 0$ to $\sim 1$. Indeed, the recovered value turns out to be $A'= 0.069$, with $1\sigma$ uncertainties of $- 0.062$ and $+ 0.69$, respectively (though the specifics here strongly depend on the chosen prior, which in our case is flat in $\log_{10}(A')$ between $10^{-2}$ and $2\pi$). This degradation in the reconstruction of the binary's phase is also reflected in the posterior for the spin parameter. Attempting to recover a secular de-phasing causes the latter to broaden significantly, and induces a bias in the recovered value of approximately $2\sigma$ with respect to the injected value. We note that this is still true for values of the accumulated dephasing that exceed $2 \pi$. We note furthermore, that attempting to recover a secular dephasing degrades the convergence of the MCMC. \newline \newline

While only based on post-Newtonian waveforms, this test case strongly indicates that the periodic perturbations of the type presented in this work are more informative and less prone to degeneracies with respect to secular dephasing, when it comes to a sophisticated parameter estimation pipeline (see Sec. \ref{sec:conclusion} for an interpretation of this result).

We are aware that a complete validation of this claim would require a more systematic study of both secular and periodic dephasing, for many sources and across an array of different waveform models, which is beyond the scope of this work. Nevertheless, the results above do suggest that waveforms of the type presented in Eqs. \ref{eq:wavemod} and \ref{eq:wavemod2} represent a strong candidate to search for EE in GW signals without necessarily degrading the estimation of the vacuum parameters of the source.

\section{Time-varying forces in astrophysical scenarios}
\label{sec:astro}
In this section we briefly explore a range of scenarios in which binary sources of GWs will be affected by time-varying forces. While we comment on the detectability of the resulting perturbations for relevant near-future GW detectors, the goal is strictly to highlight examples in which the physical processes at play naturally lead to strong fluctuations. We start by mentioning generic multipolar modifications to the gravitational potential, then delve into compact object triple systems as well as compact objects embedded in accretion discs. The discussion presented in this section is brief and mostly qualitative, though based on extrapolations from N-body simulations and hydro-dynamical simulations of gas-embedded binaries rather than purely analytical arguments.

\subsection{Generic multipole perturbations}
We apply the formalism developed in this work to a very simple, heuristic description of multi-polar modifications to the gravitational potential. We imagine the vacuum gravitational field to be perturbed by a series of additional components, arising either from the presence of background fluid, field or from modifications to GR (see e.g. \cite{2023devittiorio,2024vittorio} for a recent example). As the binary orbits in the modified potential, it will experience time varying forces with the following physical scaling:
\begin{align}
    B \sim J_{l} \frac{G M }{r^{2}}\cos \left(2l f_{\rm k }t \right),
\end{align}
where we neglect to model any projection or inclination effect, and the coefficients $J_l$ are analogous to the multi-pole moments commonly used in planetary physics. Here $l=1$ corresponds to a dipole perturbation to the gravitational potential, $l=2$ to a quadrupole perturbation, and so on. Then, the magnitude of the resulting periodic phase perturbations simply reads:
\begin{align}
    B'_{n} \approx \frac{3}{4 l^2}J_{l},
\end{align}
and the $1/r^{2}$ scaling implies:
\begin{align}
    \beta' = 0.
\end{align}
Extrapolating from the monochromatic results discussed in section \ref{sec:mono} we conclude that by employing waveforms of the type showcased by Eqs. \ref{eq:wavemod} and \ref{eq:wavemod2}, one should expect to be able to constrain completely general multipolar modifications to the gravitational potential (and therefore the spacetime metric), provided they exceed a certain typical magnitude:
\begin{align}
   J_l > J_l^{\rm det} \approx \frac{l^2}{\text{SNR}}.
\end{align}
As an example, the presence of a flattened distribution of energy density of any kind within the binary's orbit should be detectable uniquely by its induced quadrupole, provided that the total enclosed mass is of order $\sim$few/SNR. Similar considerations apply for any other multi-polar distribution.

\subsection{Dynamical and stellar triple systems}
\label{sec:alessandro}

Triple systems of compact objects are a rich laboratory of gravitational dynamics, and are being studied in great detail in the context of the dynamical formation channel for GW sources
\citep{2000ApJ...528L..17P,blaes2002,Samsing14,Liu_KozaiDamp+2015,2016PhRvD..93h4029R, 2016ApJ...824L...8R,
2016ApJ...824L...8R,2017MNRAS.464L..36A, 2017MNRAS.469.4665P,antonini2017,silsbee2017,2018PhRvD..97j3014S, 2018MNRAS.tmp.2223S,toonen2018,rodriguez2018,2018ApJ...863...68L,hoang2019,2020PhRvD.101l3010S,vignagomez2021,2022MNRAS.511.1362T,hayashi2023,pica2023, 2022MNRAS.511.1362T}.
It is known that a close-by tertiary massive body may induce several detectable imprints on GW emission from the inner binary. In this context, the most studied EEs are \eg\ Doppler modulations that arise as a consequence of Römer delays or line-of-sight accelerations, as well as the secular influence of tidal fields throughout the
inspiral \citep{2011PhRvD..83d4030Y, 2017Bonetti,2018ryu,2017PhRvD..96f3014I,
2018PhRvD..98f4012R, 2019PhRvD..99b4025C, 2019ApJ...878...75R, 2019MNRAS.488.5665W,
2020PhRvD.101f3002T, 2020PhRvD.101h3031D, 2021PhRvL.126j1105T, 2022PhRvD.105l4048S, 2023PhRvD.107d3009X, 2023arXiv231016799L, 2024MNRAS.527.8586T, 2024arXiv240305625S,2024arXiv240305625S}, {such as the secular drift caused by the von~Zeipel-Kozai-Lidov mechanism \cite{zeipel1910,koz62,lid62}}. {Here we neglect secular effects, focusing instead on the instantaneous forces experienced by the inner binary components, as they orbit in the time-varying gravitational potential induced by the tertiary.}

\begin{figure}
    \centering
    \includegraphics[width=\columnwidth]{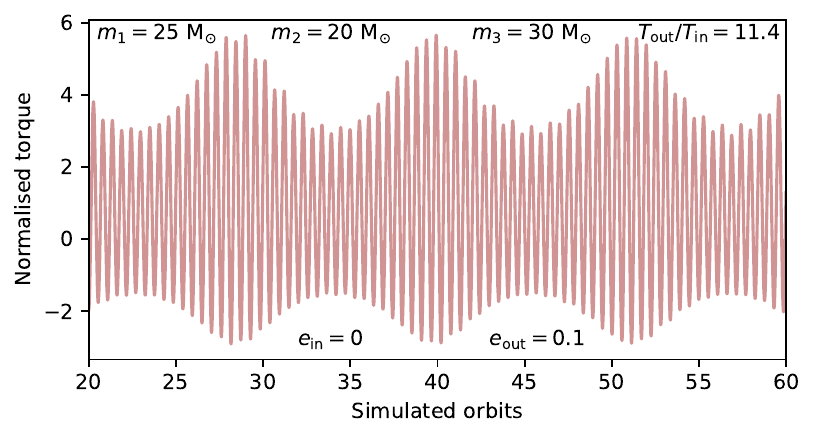}
    \includegraphics[width=\columnwidth]{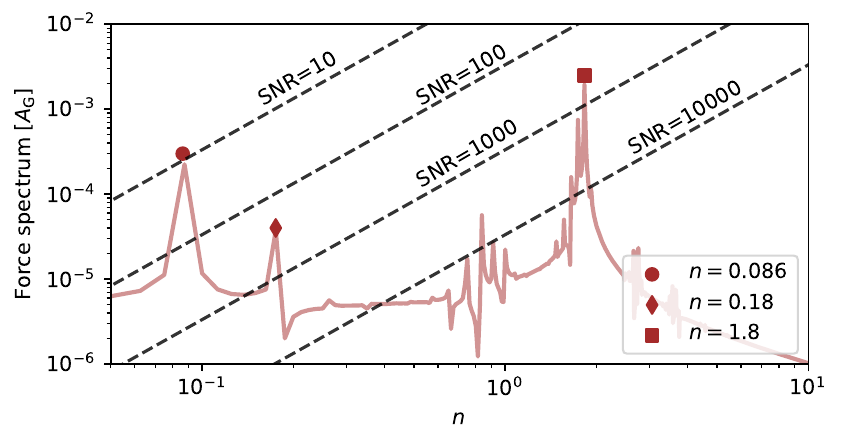}
    \caption{Top panel: Time series of the simulated torque on a binary of compact objects with masses $m_1$ and $m_2$, orbiting in the gravitational potential of a tertiary companion with mass $m_3$ (see Sec. \ref{sec:alessandro}). The outer orbit is given an eccentricity of $e_{\rm{out}}=0.1$, which is reflected in the modulation of the torques. The magnitude is normalised by its own orbit-averaged value, showcasing that strong fluctuations are present. Bottom panel: The spectrum of force Fourier-components (in the form of Eq. \ref{eq:perturbingforces}) that correspond to the simulated torque, normalised by the binary's own gravitational acceleration. We highlight three dominant force peaks at different frequencies. The dashed lines denote the approximate SNR required to detect {\textit{individual}} force peaks, according to the monochromatic criterion (see Sec. \ref{sec:detectability} and Eq. \ref{eq:monosnr}), serving as a demonstration that such peaks may be detectable for certain configurations. Note that for this realisation the $\delta$SNR of the total force spectrum is dominated by the highest peaks.}
    \label{fig:triplefag}
\end{figure}

In the top panel of Fig.~\ref{fig:triplefag} we show the time series of the instantaneous torque experienced by an inner binary due to its tertiary companion. In this specific case, the inner binary has masses of 25 and 20 $M_\odot$, while the outer companion has a mass of 30 $M_\odot$. The inner binary has a semi-major axis of $2.5\times10^{-4}$ AU, and emits GWs at 0.1 Hz. The outer binary has a semi-major axis of $1.5\times10^{-3}$ AU, yielding a period ratio of approximately 11. {The triple is coplanar, so that the von~Zeipel-Kozai-Lidov mechanism does not take place}. The simulations are performed with the regularized, direct $N$-body code \textsc{tsunami} \citep{trani2023iaus,trani2024}, which includes post-Newtonian corrections to the acceleration up to the 3.5 order.

As we can see, the torque time series presents several oscillating components at different harmonics of the inner binary's orbital frequency. The amplitude of the oscillations typically exceeds the orbit-averaged value by a factor $\sim 5$. The two major oscillatory components can by attributed to inner binary's and the outer binary's orbital motions, respectively. A more detailed view of the various components is presented in the bottom panel of Fig. \ref{fig:triplefag}, in which we express the torque time series as a periodogram in the form of Eq. \ref{eq:perturbingforces}. Note that this specific realisation of a triple system can be easily scaled to arbitrary masses and separation by approximating the potential of the tertiary as a tidal field:
\begin{align}
    B \sim \frac{G m_3}{(R + p_0)^2} - \frac{G m_3}{(R - p_0)^2} \approx \frac{4 G m_3 p_0}{R^3},
\end{align}
where $m_3$ is the tertiary's mass and $R$ is the distance of the binary to the tertiary. We observe that the torque spectrum is composed by several distinctive peaks, the most prominent of which occur at $\sim 1.8f_{\rm K}$ and $\sim 0.086 f_{\rm K}$ respectively. The exact value and relation of these two peaks encodes information about the relative orbital parameters of the binary and the tertiary. Like other secular tidal effects, time varying oscillations will therefore be stronger when the perturber is massive and close-by and are always accompanied to any other effect caused by the presence of a tertiary body. However, resolving such torque peaks would uniquely yield detailed information regarding the mass, distance and eccentricity of the latter, allowing to break the perturber mass-distance degeneracy. {As an example, the low-frequency perturbation is a consequence of the tertiary's orbit: its exact frequency scales as the ratio of periods $\propto (p_0/R)^{3/2}$, while its amplitude scales as $\propto G m_3 p_0/R^3$ and is also  mediated by the tertiary's eccentricity}.

\subsection{Massive black holes in circumbinary discs}
\label{sec:chris}
\begin{figure}
    \centering
    \includegraphics[width=\columnwidth]{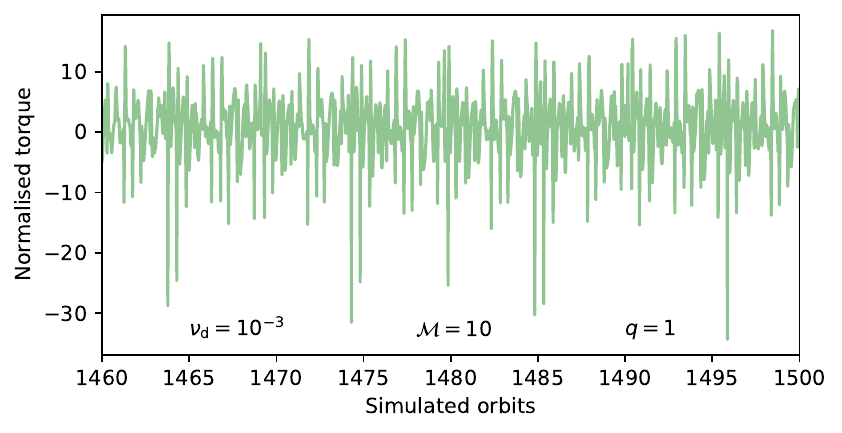}
    \includegraphics[width=\columnwidth]{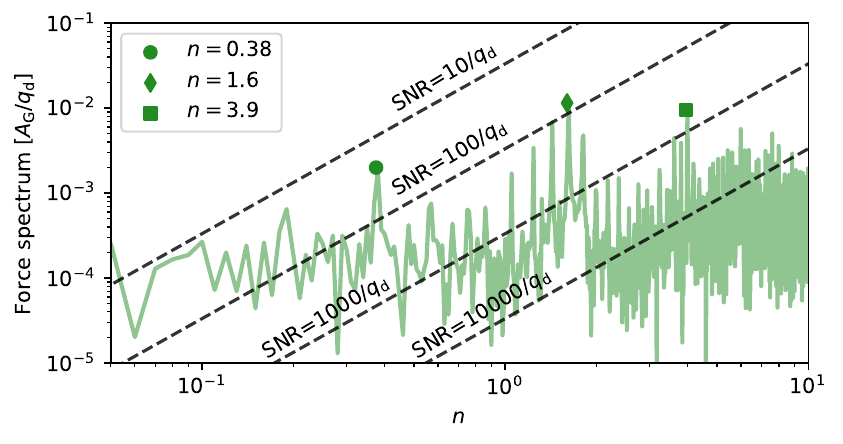}
    \caption{Top panel: Time series of the simulated hydro-dynamical torque on an equal-mass SMBH binary, orbiting in a viscous accretion disc (see Sec. \ref{sec:chris}). The magnitude is normalised by its own orbit-averaged value, showcasing that strong fluctuations are present. Bottom panel: The spectrum of force Fourier-components (in the form of Eq. \ref{eq:perturbingforces}) that arise from the simulated torque, normalised by the binary's own gravitational acceleration. We highlight three dominant force peaks at different frequencies. The dashed lines denote the approximate SNR required to detect {\textit{individual}} force peaks, according to the monochromatic criterion (see Sec. \ref{sec:detectability} and Eq. \ref{eq:monosnr}). Here the SNR lines are scaled with the disc's scale-mass to binary-mass-ratio, $q_{\rm d}$, defined in Eq. \ref{eq:qd}. This example  serves as a demonstration that such peaks may be detectable for certain binary-disc configurations. Note that for this realisation of a torque spectrum, the total $\delta$SNR of the resulting phase perturbation is significantly larger than what is suggested by the largest individual peaks.
    }
    \label{fig:tiede-torques}
\end{figure}

%\ct{Maybe makes sense to introduce E/IMRIs first? e.g. "majority of work on gas-embedded LISA sources has been on high-mass difference sources and has focused on the secular accumulation of de-phasing... These focus on the time-averaged forces and torques on the binary orbit and wash away variability information on dynamical times..."}

%\zh{I think PTAs should be mentioned here, since gas can have a huge effect for the wider PTA sources -- the secular changes can be of order unity, but if there are periodic fluctuations for them, like in Fig.9, could that be detectable too?}

% \begin{itemize}
%     \item gas-embedded SMBHBs : why care about this?
%         \subitem- many observed sources expected to undergo this phase
%         \subitem- may be important mechanism to shepherd into GW regime
%     \item Similar-mass systems primarily studied with hydro simulations
%         \subitem- periodicity in accretion rates a primary focus
%         \subitem- torques only considered over long timescales, but also demonstrate periodic behavior -- this has not previously been documented/analyzed  (I checked this once but should double check)
%     \item Recent works have also suggested discs stay attached to inspiralling binaries significantly into the LISA band (and may never decouple at all)
%     \item Periodic forcing from circumbinary material may be detectable in individual sources
%     \item Analyze simulation data from \texttt{Sailfish} with setup like SB Code-Comp, measure torques, and fourier decompose
% \end{itemize}

The loudest and highest-quality GW sources of future space-based detectors will come from comparable-mass, massive black holes coalescing in the m-Hz band.
Such binaries are expected to form following galaxy mergers, which are known to efficiently funnel material into the resulting galactic nucleus \citep{Barnes_Hernquist_1996, Springel:2005}. Therefore, many SMBH binaries are expected to undergo gas-embedded accretion phases which may play an important role in shepherding them from initially large separations down to GW-dominated regimes \citep{Begel:Blan:Rees:1980, Milosavljevic:2003, 2009haiman}. Modern studies have also suggested that circum-binary material can remain coupled to the binary significantly in to a GW-dominated regime \citep{Farris_dec+2015, Dittmann:Decoupling:2023, Krauth:2023, Avara:3DMHD:2023, Cocchiararo:2024} and that secular gas-induced changes to the binary's orbital elements could feasibly be detected by LISA \citep{garg2022, 2024garg, 2023Tiede}. However, the study of the time-variability in such forces has largely been omitted.

The dynamics of gas-embedded comparable-mass systems are highly non-trivial and have primarily been studied numerically \citep[for a recent review see][]{LaiMunoz:Review:2022}. A primary focus of such investigations has been the periodicity of the accretion rate onto the binary (which is not always dominated by the binary's orbital frequency) and its applications to electromagnetic searches for compact massive binaries \citep[e.g.][]{D'Orazio:binlite:2024,2024franchini}. In \cite{Roedig_Trqs+2012} a spectral analysis of the torques in such systems was performed, but nearly all recent analyses of the gas forcing and resultant change in orbital elements have focused on the time-averaged, secular evolution.

In the top panel of Fig. \ref{fig:tiede-torques} we illustrate a raw time-series of the torques experienced by an equal-mass binary fixed on a circular orbit over 40 periods, {which arise from both gravitational forces and gas accretion and vary quasi-periodically at multiple frequencies}. We see that the resultant torques in a quasi-steady configuration are in fact highly variable, with features beyond those solely at the binary orbital frequency. Here the magnitude of the torque shows fluctuations by a factor of order $\sim 30$ with respect to its orbit-averaged value. The presented data is taken from a 2D, isothermal-hydrodynamics simulation performed with \texttt{Sailfish} \citep[c.f.][]{Westernacher-Schneider:2022}.
The setup is exactly that presented in \cite{SB-CodeComp:2024} for a thin, co-planar disc of characteristic scale-height $h/r \approx \mathcal{M}^{-1} = 0.1$ and global kinematic viscosity $\nu = \nu_{\rm d} a^2\Omega_b$ with $\nu_{\rm d} = 10^{-3}$ (roughly equivalent to a turbulo-viscous parameter $\alpha = 0.1$ at $r = a$; \cite{1973shakura}).

The bottom panel of Fig. \ref{fig:tiede-torques} shows a periodogram of the components of the torque. In comparison to three-body systems, hydro-dynamical torques are significantly more complex and spread over a wide range of harmonics. Therefore, we expect the $\delta$SNR of the resulting phase perturbations to be significantly larger than what individual torque component peaks may suggest (following Eq. \ref{eq:quadrature}). Nevertheless, we can identify several peaks at approximately $\sim 0.38 f_{\rm K} $, $\sim 1.6 f_{\rm{K}}$ and $\sim 3.9 f_{\rm K}$.
%, which can be associated with different features in the morphology of the circum-binary disc. 
Therefore, a high-SNR observation of GWs from such a system would in principle allow one to reconstruct features in the flow of gas around in-spiralling black holes. {We note that the numerical tests illustrated in section \ref{sec:detectability} strongly suggest that periodic perturbations will be orthogonal the secular de-phasing arising from accretion or gravitational torques, which necessarily accompany them. The latter have been studied in detail in e.g. \cite{garg2022,2023zwick} and \cite{2023Tiede}. The relative importance of periodic versus secular perturbations will strongly depend on the properties of the accretion disc. A joint detection of secular dephasing and periodic perturbations would be an unprecedented probe of the disc's structure, and will be the focus of future work.} We also note that {our qualitative statements} should hold for any type of gas-embedded binary, including stellar-mass binaries in accretion discs \citep[\eg][]{Antoni:2019, LiLai:2022, Dempsey3D:2022, rowan2023, 2023whitehead, DittmannDempsey:2024}.

The results displayed in the bottom panel of Fig. \ref{fig:tiede-torques} scale with a dimensionless parameter $q_{\rm d}$:
\begin{align}
    \label{eq:qd}
    q_{\rm d} = \frac{1}{3\pi \nu_{\rm d}} \frac{\dot{M}}{M}(2 \pi f_{\rm K})^{-1},
\end{align}
where $\dot{M}$ is the accretion rate onto the binary. The parameter $q_{\rm d}$ can be roughly understood as the ratio of the disc's scale mass and the binary total mass, which depends on the binary separation \citep[compare with \eg][]{2023Tiede}. For realistic accretion discs this parameter will be small when sources enter the m-Hz band, of order $\sim 10^{-5}$ to few $\sim 10^{-1}$, depending on the total binary mass and the properties of the disc. 
% {One paragraph on fluctuations increasing for larger mach numbers, which are more realistic? You got this chris save me <3 <3 }
% \textcolor{violet}{Here you go bb ;)}

We also note that while the above analysis and the majority of numerical work on this problem has focused on discs of characteristic scale-height $h/r = 0.1$, steady-state modeling suggests that discs around such massive central objects ought to be much thinner with $h/r \sim \mathcal{O}(10^{-2} - 10^{-3})$ \citep{AccretionPower, sirko2003, thompson2005}.
A series of recent studies have documented how the binary evolution and accretion morphology are sensitive to the disc scale height \citep{tiede2020, Dittmann:2022, Penzlin:2022, Franchini:2022}.
Of primary note, both the net gravitational torque on the binary and the magnitude of torque fluctuations have been observed to grow meaningfully with decreasing $h/r$ \citep{tiede2020, Derdzinksi:2021}.
Therefore, the results in Fig. \ref{fig:tiede-torques} might be regarded as conservative if thin, radiatively efficient discs around massive binaries resemble those around their single-BH AGN-counterparts\footnote{But see also \cite{Cocchiararo:2024} who find that binary heating of the inner disc can sustain thicker discs akin to the $h/r=0.1$ case presented here.}.

% \dd{Mention also the torque spectrum in \citep{Roedig_Trqs+2012}}
%https://arxiv.org/pdf/1202.6063.pdf Fig 7

\subsection{EMRIs and IMRIs in accretion discs}
\label{sec:andrea}
Extreme and intermediate mass-ratio inspirals (E/IMRI) consist in a stellar-mass BH orbiting a central SMBH. Such systems may form in the accretion discs of active galactic nuclei at potentially high rates, via the process of BH-accretion and disc interaction \citep{tagawa2020,fabj2020,2021wet,delaurentis2023,rowan2023} and/or in-situ formation of embedded stars \citep{2007levin,2022derdz}.
For this particular type of GW source, characterising the environmental properties is critical, as a misinterpretation of EEs can interfere with precision measurements of BH spin and of the central metric's multipolar structure \citep{2014barausse,2022speri}. 

\begin{figure}
    \centering
    \includegraphics[width=\columnwidth]{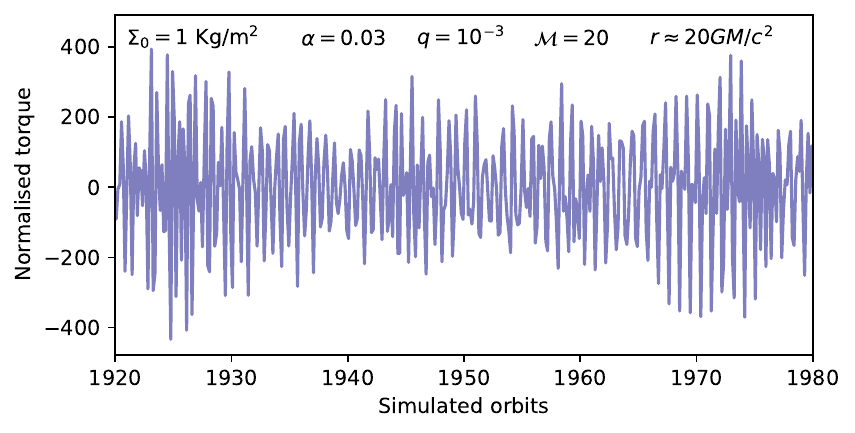}
    \includegraphics[width=\columnwidth]{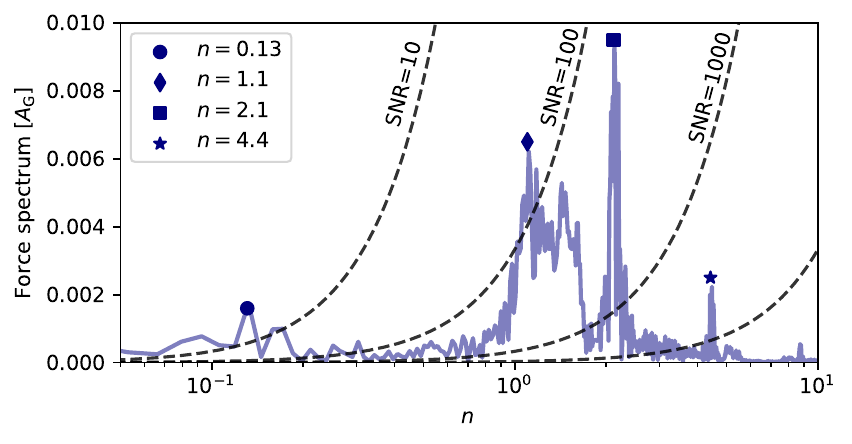}
    \caption{Top panel: Time series of the simulated hydro-dynamical torque on an intermediate mass-ratio inspiral ($q=10^{-3}$) in a viscous accretion disc (see Sec. \ref{sec:andrea}). The magnitude is normalised by its own orbit-averaged value, showcasing how fluctuations dominate the short time-scale dynamics of the system. Bottom panel: The spectrum of force Fourier-components (in the form of Eq. \ref{eq:perturbingforces}) that arise from the simulated torque. We highlight three dominant force peaks at different frequencies, related to features of the gas flow around the perturber. The dashed lines denote the approximate SNR required to detect {\textit{individual}} force peaks, according to the monochromatic criterion (see Sec. \ref{sec:detectability} and Eq. \ref{eq:monosnr}), serving as a demonstration that such peaks may be detectable for high SNR IMRIs and EMRIs. Note that for this realisation the total $\delta$SNR of the total force spectrum is be significantly larger than what is suggested by the individual peaks.
    }
    \label{fig:imri_sim_andrea}
\end{figure}

Simulations exploring E/IMRI coalescence in accretion discs find that the gas exerts torques with fluctuating magnitude on the embedded BH, which roughly averages out to match the expectations of linear perturbation theory and planetary migration over many orbits \citep{2002ApJ...565.1257T,1980ApJ...241..425G,2021andrea}. In Fig. \ref{fig:imri_sim_andrea}, we show a time-series of the instantaneous torque experienced by an IMRI embedded in a steady-state, iso-thermal and viscous accretion disc. The data are taken from un-smoothed results of the simulations detailed in \cite{2021andrea}, i.e. 2D hydro-dynamical Newtonian simulation performed with the code DISCO \citep{2016ApJS..226....2D}, in which an IMRI slowly in-spirals on a prescribed, quasi-circular orbit. The parameters for the simulation are a binary mass-ratio $q=10^{-3}$, a disc alpha viscosity parameter $\alpha=0.03$, and a uniform disc scale height $h/r \approx 0.03$. Here we additionally choose a conservative surface density scale for the disc of $0.1$ g/cm$^{2}$ \citep[compare with \eg,][]{2021andrea}, which already suffices to produce very strong effects. Note that both the secular torques and the fluctuations scale linearly with the disc's scale density.

During the in-spiral, the IMRI experiences torque fluctuations on various timescales comparable to the orbital timescale, which typically exceed their orbit-averaged values by a factor of several hundred. Such strong fluctuations are interpreted to be a consequence of gravitational forces arising from the asymmetry in the mini-disc around the secondary BH. In the bottom panel of Fig. \ref{fig:imri_sim_andrea}, we showcase a periodogram of the acceleration suffered by the embedded IMRI, normalised by the gravitational acceleration due to the central object. We see that the force spectrum is rich with fluctuations at multiple harmonics, though presents some distinct peaks at $\sim 0.13 f_{\rm K}$, $\sim 1.1 f_{\rm K}$ and $\sim 2.1 f_{\rm K}$. Detecting such components in an IMRI signal would provide information regarding the gas dynamics around the embedded BH, as well as the global response of the disc to a small perturber. This is particularly useful, as GW distortions are probes of the nearby gas mass distribution, while all of our current information from EM observations only probes the surface of such optically thick discs. Similarly to the circumbinary disc case, the total torque spectrum on an embedded IMRI  will produce phase perturbations with a significantly larger $\delta$SNR than the highest peaks in the periodogram would suggest (according to Eq. \ref{eq:quadrature}). We note that at later stages of the in-spiral, the magnitude of fluctuations grows (albeit remains weak compared to the dominance of GW induced evolution), while the periodicity remains correlated with the binary orbital period. Note also that the torque fluctuations for this system are enhanced by the imposed GW in-spiral, an effect which is not explored in our demonstrated example of MBHB interaction (Fig. \ref{fig:tiede-torques}).

Aside from the nonlinear torque evolution discovered in laminar accretion flows, we comment here that accretion discs are also expected to be magnetized and turbulent \citep{1991ApJ...376..214B}. This may cause stochastic migration for lower mass BHs (EMRIs), which would result in additional periodic torque fluctuations. 
For an example, we refer the reader to the simulation results of \cite{2024WuChenLin}, which explore migration in turbulent accretion disc models. In short, it is expected that gas-embedded IMRIs and EMRIs will be subject to a wide range of physical processes that produce torque fluctuations in addition to the secular energy and angular momentum fluxes (see also \cite{2022zwick} for a thorough discussion).

%%%%

\section{Conclusions}
\label{sec:conclusion}
In this work, we have characterised aspects of the motion of perturbed circular binaries (Sec. \ref{sec:GWfrombin}) and the resulting GW emission, focusing specifically on time-varying forces that do not cause any secular evolution in the orbital elements. We constructed a waveform model, which we later use to produce detectability criteria for the resulting periodic perturbations in the GW phase (Sec. \ref{sec:detmethods} and \ref{sec:detectability}). Finally, we discussed a range of astrophysical scenarios in which such time-varying forces occur naturally, and often at significantly larger amplitudes that secular effects (Sec. \ref{sec:astro}).

A major unanswered question remains: Are sufficiently large periodic phase perturbations realised in nature, such that they may indeed be extracted from realistic GW signals with near-future instruments? In Sec. \ref{sec:astro}, we have shown how many astrophysical scenarios will naturally produce strong force fluctuations that greatly exceed the magnitude of orbit-averaged forces, basing our arguments on the results of N-body and hydro-dynamical simulations. Our examples show that the potential exists to detect such perturbations in astrophysical sources that are expected for near future detectors. We nevertheless stress that answering the aforementioned question requires a thorough estimate of detection rates. Beyond the intrinsic modelling uncertainties for the perturbations in question, the latter would have to entail the following elements:
\begin{itemize}
    \item Population synthesis models for each of the GW source types and GW detectors mentioned in this work.
    \item A complete data analysis pipeline that takes all GW parameters into account, including spins, various orientation angles and potentially the additional GW perturbations discussed in Sec. \ref{sec:GWfrombin}.
    \item Accounting for the full force spectra rather than individual harmonics. This is in particular relevant for the complex forces expected to act on gas-embedded sources.
    \item  A thorough analysis of the odds of periodic perturbations being related to astrophysical EEs rather than some unknown sources of noise.
\end{itemize}
While such an analysis is beyond the scope of this paper, we do stress that the detection of even a single binary source that is clearly perturbed by periodic EEs would be momentous, as it would represent unequivocal and informative evidence of the interaction of a binary in the GW dominated phase with its surroundings. Additionally, the results of this work highlight the potential of considering periodic perturbations in GWs, in comparison with \eg\, secular ones. To reiterate the argument: In Sec. \ref{sec:degen} we have discussed how a hypothetically constant secular de-phasing, i.e. a phase offset, suffers from being entirely degenerate with the source's initial phase (for values below $2 \pi$). In other words, it is only the rate of variation of de-phasing that is informative, rather than the magnitude of de-phasing itself. Crucially, the periodic phase perturbations of the type analysed in this work essentially \textit{maximise the rate of variation} of the perturbation in the GW phase by construction. They are therefore maximally detectable as an EE, given a typical amplitude. Furthermore, they are less prone to result in ulterior degeneracies, or in the degradation of the recovery of vacuum parameters (see Fig. \ref{fig:mcmc2} and Sec. \ref{sec:degen}). Indeed, the question of when the accumulation of secular de-phasing over long timescales out-competes the  advantages of periodic de-phasing also requires more investigation.

In the meantime, we suggest that the latter be included in future studies of EEs by means of Eqs. \ref{eq:wavemod} and \ref{eq:wavemod2} (or any other extension), in addition to secular de-phasing prescriptions.

\begin{acknowledgments}
L.Z. acknowledges support from  ERC Starting Grant No. 121817–BlackHoleMergs.
A.A.T. acknowledges support from the European Union’s Horizon 2020 and Horizon Europe research and innovation programs under the Marie Sk\l{}odowska-Curie grant agreements No.~847523 and 101103134.
D.J.D. and C.T. received support from the Danish Independent Research Fund through Sapere Aude Starting grant No. 121587.
Z.H. acknowledges support from NSF grant AST-2006176 and NASA grants 80NSSC22K0822 and 80NSSC24K0440.
AD acknowledges support from NSF grant AST-2319441.
The authors thank the referee for many useful comments.
\end{acknowledgments}

% The \nocite command causes all entries in a bibliography to be printed out
% whether or not they are actually referenced in the text. This is appropriate
% for the sample file to show the different styles of references, but authors
% most likely will not want to use it.
%\nocite{*}

\bibliography{main}
\bibliographystyle{apj}

% Produces the bibliography via BibTeX.

\end{document}